\documentclass[conference]{IEEEtran}

\usepackage{cite}

\ifCLASSINFOpdf
   \usepackage[pdftex]{graphicx}
   \graphicspath{{../pdf/}{../jpeg/}}
   \DeclareGraphicsExtensions{.pdf,.jpeg,.png}
\else
\fi

\usepackage[cmex10]{amsmath}

\usepackage{array}

\usepackage[table,x11names]{xcolor}
\usepackage{slashbox}
\usepackage{xfrac}
\usepackage{pgfplots}
\usepackage{tikz}
\usepackage{algpseudocode}
\usepackage{algorithm}
\usepackage{amsfonts}
\usepackage{amssymb}
\usepackage{amsbsy}
\usepackage{amsthm, mathrsfs}
\usepackage{longtable}
\usepackage{mathtools}
\usepackage{array} 
\usepackage{cite}
\usepackage{dsfont}
\usepackage{graphicx}
\usepackage{subcaption}
\usepackage{caption}
\usepackage{tabularx}

\usepackage{wrapfig}

\usepackage{multirow}
\usepackage{rotating}

\theoremstyle{definition}
\newtheorem{definition}{Definition}

\usepackage{paralist}

\usepackage{fancyhdr}
\usepackage{kantlipsum}
\fancypagestyle{plain}{
  \fancyhf{}
  \fancyhead[C]{Conference on \LaTeX}     
  \fancyfoot[L]{This is a notice}

}
\usepackage{eso-pic}

\begin{document}
%
\title{Privacy Preserving Identification Using \\Sparse Approximation with Ambiguization}


\author{\IEEEauthorblockN{Behrooz~Razeghi, Slava~Voloshynovskiy, Dimche~Kostadinov and Olga~Taran}
\IEEEauthorblockA{Stochastic Information Processing Group, Department of Computer Science, University of Geneva, Switzerland\\
 \{\texttt{behrooz.razeghi, svolos, dimche.kostadinov, olga.taran}\}@\texttt{unige.ch}}}
%




\maketitle

\begin{abstract}
In this paper, we consider a privacy preserving encoding framework for identification applications covering biometrics, physical object security and the Internet of Things (IoT). The proposed framework is based on a sparsifying transform, which consists of a \textit{trained linear map}, an \textit{element-wise nonlinearity}, and \textit{privacy amplification}. The sparsifying transform and privacy amplification are not symmetric for the data owner and data user. We demonstrate that the proposed approach is closely related to sparse ternary codes (STC), a recent information-theoretic concept proposed for fast approximate nearest neighbor (ANN) search in high dimensional feature spaces that being machine learning in nature also offers significant benefits in comparison to sparse approximation and binary embedding approaches. 
We demonstrate that the privacy of the database outsourced to a server as well as the privacy of the data user are preserved at a low computational cost, storage and communication burdens.
\end{abstract}


%
\IEEEpeerreviewmaketitle

\vspace{-3pt}


\section{Introduction}

\vspace{-3pt}

\subsection{Identification and ANN Search}

\vspace{-3pt}

Many modern applications such as biometrics, digital physical object security and data generated by connected objects in the IoT require privacy preserving identification of a query with respect to a given dataset. Practically, the identification problem is based on an ANN search when a list of indices corresponding to the NN items is returned. At the final refinement stage, the list can be refined in a private setting and a single index is declared as the identified one. The identification problem faces the curse of dimensionality. For this reason, the exact identification is replaced by a search of list of closest items, i.e., one tries to trade-off the accuracy of identification by the search complexity. In recent years, many methods providing efficient ANN solutions for multi-billion entry datasets were proposed and we named some of them without pretending to be exhaustive in our overview \cite{Sohrab_WIFS2016, mathon2013secure, rane2012attribute}. 

\vspace{-3pt}

\subsection{Search~in~Privacy~Preserving~Settings:~Main~Considerations}

\vspace{-3pt}

Due to the massive amount of data, modern distributed storage and computing facilities, many ANN problems are considered in a setting where the data user outsources his datasets by applying the corresponding protection measures to third parties (servers) possessing powerful storage, communications and computing facilities. The need for data protection comes from many perspectives related to the cost of data collection, data as a "product" that represents a great value in the era of machine learning, which can be used to train and prune new and existing machine learning tools. Moreover, the server might want to discover some hidden relationships in the data. Finally, the non-renewable nature of some features such as for example biometrics, which being disclosed once, do not represent any more a value for the related security applications. The data users (clients), who by hypothesis poses some query, related to those stored in the databases of the data owners, wish to identify them by obtaining an index/indices of the closest items in the data owner datasets. However, the client does not want to disclose his query to the server completely for privacy reasons. Therefore, it is assumed that both the data owner and clients attempt at protecting their data from server side analysis, which is assumed to be honest but curious. 

\vspace{-3pt}

More particularly, the server might attempt: (a) to try to find relationships between entries in the database of the data owner, (b) reconstruct an individual entry of the data owner database or a common representatives or centroids for the clusters, (c) reconstruct the query of clients and (d) cluster multiple queries from the same client or from multiple ones thus establishing the group interests based on the similarity of probes. Curious clients might be also interested to discover more information about the structure of database by exploring the NN around one or multiple probes. Additionally, one can envision collaborative clients who might aggregate the results for multiple queries.\IEEEpubidadjcol

\vspace{-2pt}

\begin{figure}[!t]
\centering
\includegraphics[scale=0.45]{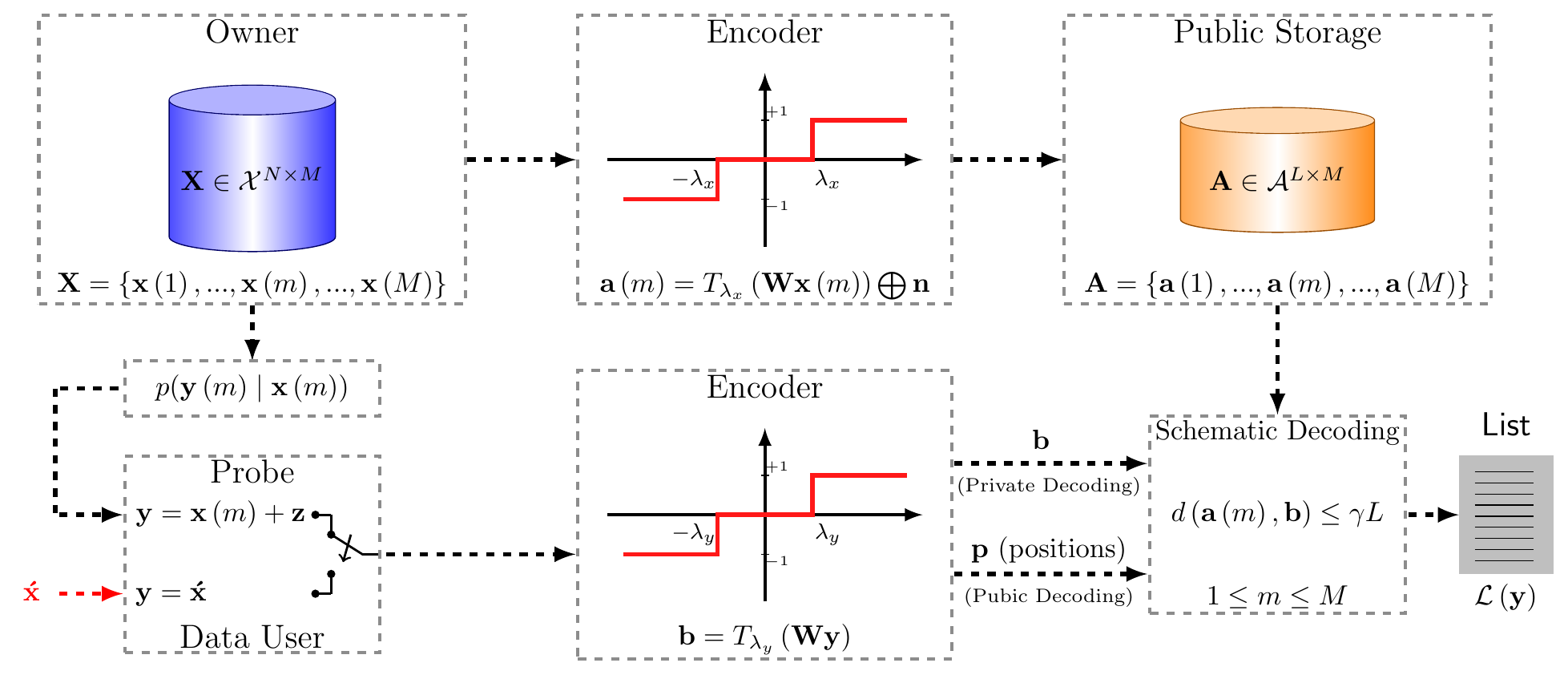}
\vspace{-15pt}
\caption{Block diagram of the proposed model.}
\vspace{-20pt}
 \label{Fig:Model}
\end{figure}

\vspace{-6pt}


\subsection{Our contribution}

\vspace{-4pt}

This paper presents a new privacy preserving strategy to the ANN search problem based on a recently proposed concept of Sparse Ternary Codes \cite{Sohrab_WIFS2016, Sohrab_ISIT2016}. Our main contribution consists in a novel formulation of database protection based on a STC representation with the addition of ambiguization noise to prevent database analysis and a novel formulation of the query mechanism based on sending the positions of non-zero components in the query sparse representation along with the fraction of ambiguous positions to prevent the reconstruction from query and collaborative processing of queries. Our approach differs from existing privacy protection based on quantized embedding \cite{boufounos2011secure, mathon2013secure} and attribute based \cite{iscen2016group} techniques.
It has strong information-theoretic foundations and demonstrated higher coding gain with respect to binarized embedding methods \cite{Sohrab_ISIT2016}.

\vspace{-4pt}

Our setup is quite generic and we assume that as an input we might have raw data, extracted features using any known hand crafted methods, aggregated local descriptors based on BoW, FV, VLAD \cite{jegou2009burstiness, perronnin2007fisher, jegou2010aggregating}, etc, or intermediate or last layers of deep nets \cite{babenko2014neural}.
Before we provide the input to the server we preprocess it by passing it via the sparsifying transform and thresholding it thus producing STC.\footnote{ It should be noted that even higher flexibility is possible for multilevel quantization.} The next stage is an ambiguation that consists in the addition of selective noise components to the zero positions in the sparse data representation (Fig.~\ref{Fig:Model}).
As a sparsifying transform we use an efficient learning method based on the sparsified Procrustes problem formulation \cite{ravishankar2013closed} in contrast to a fixed transform used in \cite{Sohrab_WIFS2016, Sohrab_ISIT2016}.
The resulting data are stored on the \textit{public} server. The client, who possesses some query, sends a set of non-zero positions after sparsification of the query to the server and expects that the server will return a set of corresponding indices corresponding to positive and negative non-zero components. To ambiguise the server, the data user also adds a certain portion of false positions to the query and keeps their indices. The search complexity of the server is logarithmic in terms of the number of stored items times the number of non-zero positions in the sparsified query. This also determines the returned file size to the client. Due to the efficient sparse representation, the amount of returned data is minimized while preserving the  efficiency of the data structure representation. At the final stage, the data user disregards the lists corresponding to false positions and aggregates the remaining lists for the corresponding positions producing the final ANN list.

\vspace{-8pt}

\subsection{Notation and Definitions}

\vspace{-4pt}

Throughout this paper, superscript $(.)^T$ stands for the transpose and $(.)^\dagger$ stands for pseudo-inverse. Vectors and matrices are denoted by boldface lower-case ($\mathbf{x}$) and upper-case ($\mathbf{X}$) letters, respectively. We consider the same notation for a random vector $\mathbf{x}$ and its realization. The difference should be clear from the context. $x_i$ denotes the $i$-th entry of vector $\mathbf{x}$. For a matrix $\mathbf{X}$, $\mathbf{x}{\left(  j \right)}$ denotes the $j$-th column of $\mathbf{X}$.
 Also, we use the notations $\left[ N \right]$ for the set $\{ 1, 2, 3, ..., N\}$ and $\mathrm{card} \left(\mathcal{S}\right)$ for the cardinality of a set $\mathcal{S}$.  
\vspace{-3pt} 

\theoremstyle{definition}
\begin{definition}{}
The support of a vector $\mathbf{x} \in \mathbb{R}^N$ is the index set of its nonzero entries, i.e.,
$\mathrm{supp} \left( \mathbf{x} \right) \coloneqq \{ n \in \left[ N \right] : x_n \neq 0 \}.$
\end{definition}

\vspace{-5pt}

\theoremstyle{definition}
\begin{definition}{}
Given a subset $\mathcal{E}$ of a space $\mathcal{X}$, the indicator function $ \mathds{1}_{\mathcal{E}} \, : \, \mathcal{X} \rightarrow \mathbb{R}$ is defined by setting $\mathds{1}_{\mathcal{E}} \left(x \right)$ equal to $1$ for $x \in \mathcal{E}$ and equal to $0$ for $x \notin \mathcal{E}$. 
\end{definition}

\vspace{-8pt}

\subsection{Outline of the Paper}
\vspace{-2pt}
 
The remainder of the paper is organized as follows. In Section \ref{Sec:II}, the identification problem is defined. We then propose our framework in Section \ref{Sec:III}. Section \ref{Sec:IV} evaluates the privacy imposed by our framework. Finally, conclusions are drawn in Section \ref{Sec:V}.

\vspace{-8pt} 

\section{Identification Problem}\label{Sec:II}

\vspace{-3pt}

Consider that $M$ (raw) feature vectors are stored in the database $\mathbf{X} =  \left[ 
\begin{array}{ccccc}
\mathbf{x}{\left(1\right)} \! &  \cdots  &  \! \mathbf{x}{\left(m\right)} \! &    \cdots  \! & \mathbf{x}{\left(M\right)}
\end{array}
  \! \right]$, 
where each feature vector $\mathbf{x}{\left(m\right)}, m \in \left[M\right]$ from a set $\mathcal{X} \subset {\mathbb{R}}^{N}$ is a random vector with distribution 
$p\left( \mathbf{x} \right)$ and bounded variance $\sigma^2_{\mathbf{x}}$.

The query (probe) $\mathbf{y}{\left(m\right)}  \in {\mathbb{R}}^N$ is the noisy version of $\mathbf{x}{\left(m\right)}$, i.e., $\mathbf{y}{\left(m\right)} = \mathbf{x}{\left(m\right)} + \mathbf{z}$, where we assume $\mathbf{z} \in {\mathbb{R}}^N$ is a Gaussian noise vector with distribution $\mathcal{N} \left( \mathbf{0}, \mathbf{\Sigma}_{\mathbf{z}} =  \sigma^2_{\mathbf{z}} \mathbf{I}_N \right)$. 
Provided that the additive noise $\mathbf{z}$ is independent of data, for large $N$ the law of large numbers states that 
$\frac{1}{N} { \| \mathbf{y}{\left(m\right)} - \mathbf{x}{\left(m\right)} \| }_2^2 \!    \rightarrow  \frac{1}{N}  \mathbb{E}   \big[ { \mathbf{z} }^{T}  \mathbf{z} \big]     \simeq    \sigma^2 _{\mathbf{z} }, \, \forall m \in \left[ M \right]$. 
Therefore, for an enrollment vector $\mathbf{x}{\left(m\right)}$ stored at  database with high probability query vector $\mathbf{y}{\left(m\right)}$ will be in an $N$-dimensional sphere of radius $\sqrt{N} \sigma_{\mathbf{z}} $ and centered at $\mathbf{x}{\left(m\right)}$.

\vspace{-6pt}
 
 
\section{Proposed Framework}\label{Sec:III}

\vspace{-3pt}
\subsection{Sparse Representation and Encoding}
\vspace{-2pt}

The intrinsic information content of feature vectors, like other real-world signals, is much smaller than their lengths. So we are interested in approximating them by sparse vectors. 
Many signals $\mathbf{x}{\left(m\right)} \! \in \! {\mathbb{R}}^N$ may be represented as a linear combination of a small number of columns (words) from a frame\footnote{A set of $L$ orthonormal vectors with vector dimensionality $N$ equal to $L$ is said to form a basis set for that vector space.  A frame of an inner product space is a generalization of a basis of a vector space to sets that may be linearly dependent.} (dictionary) $\mathbf{D} \in {\mathbb{R}}^{N \times L}$ according to the sparse approximation model. Therefore, $\mathbf{x}{\left(m\right)} \! = \mathbf{D} \mathbf{a}{\left(m\right)} + \mathbf{e}_{\mathbf{x}} $, where $\mathbf{e}_{\mathbf{x}} \in {\mathbb{R}}^{N}$ is approximation error, $\mathbf{a}{\left(m\right)}  \in {\mathbb{R}}^{L}$ is sparse with ${\| \mathbf{a}{\left(m\right)}  \| }_0 \ll L$, in which ${\| \mathbf{a}{\left(m\right)} \| }_0 \!  \coloneqq  \mathrm{card} \! \left( \mathrm{supp} \left( \mathbf{a}{\left(m\right)} \right) \right)$. 
The general problem formulation of this sparse coding problem for a fixed dictionary can be expressed~as: \vspace{-3pt}
\begin{equation}
\mathbf{\widehat{a}}{(m)} \!=\! \mathop{\arg \min}_{\mathbf{a}{\left(m\right)} \in \mathcal{A}^L} { \| \mathbf{x}{\left(m\right)}  \! - \! \mathbf{D} \mathbf{a}{\left(m\right)} \|}_2^2 + \lambda \Omega \left(\mathbf{a}{\left(m\right)} \right), \! \forall m \in \! \left[ M \right],
\end{equation}\vspace{-2pt}%
where $\mathcal{A}$ is an alphabet of sparse representations, $\lambda$ is a regularization parameter and $\Omega \left( . \right)$ is sparsity-inducing regularization function. However, the sparse coding with respect to $\mathbf{a} \left(m\right)$ is an inverse problem in nature and known to be NP-hard. This problem can be solved approximately by greedy or relaxation algorithms, but these only provide the correct solution under certain conditions, which are often violated in real world applications. Furthermore, they are computationally expensive in practice, particularly for large-scale problems.

\vspace{-2pt}

In this paper, we use a transform model \cite{ravishankar2013learning} for the sparse representation of vectors at enrollment and identification phases. This model suggests that a feature vector $\mathbf{x}{\left(m\right)} \in \mathbb{R}^N$ is approximately sparsifiable using a transform $\mathbf{W} \in \mathbb{R}^{L \times N}$, that is $\mathbf{W} \mathbf{x}{\left(m\right)} = \mathbf{a}{\left(m\right)} + \mathbf{e}_{\mathbf{a}}$, where $\mathbf{a}{\left(m\right)} \in \mathbb{R}^L$ is sparse, i.e., ${\| \mathbf{a}{\left(m\right)} \|}_{0} \ll L$, and $\mathbf{e}_{\mathbf{a}} \in \mathbb{R}^L$ is the representation error of the feature vector or residual in the \textit{transform domain}. The sparse coding problem for this model is a direct constraint projection problem. The general problem formulation, also known as sparse approximation, is as follows: \vspace{-3pt}
\begin{equation}\label{l_0-optimalsolution} 
\mathbf{\widehat{a}}{(m)} \!=\!  \mathop{\arg \min}_{\mathbf{a}{\left(m\right)} \in \mathcal{A}^L} { \| \mathbf{W} \mathbf{x}{\left(m\right)} \!  -\!  \mathbf{a}{\left(m\right)} \|}_2^2 + \lambda \Omega \left(\mathbf{a}{\left(m\right)} \right)\! , \forall m \in \! \left[ M \right].
\end{equation} 

 \vspace{-5pt}

Note that the residual and therefore the sparse representation $\mathbf{a}{\left(m\right)}$ is not constrained to lie in the range space of $\mathbf{W}$. 
Furthermore, since it is a direct problem, for any of the two important regularizers $\Omega \left( . \right) =  {\| . \|}_0$ or $\Omega \left( . \right) =  {\| . \|}_1$ has closed-form solutions. In this paper, we consider $l_0$-``norm'' as our sparsity-inducing penalty. In this case, the solution $\mathbf{\widehat{a}}{(m)}$ is obtained exactly by hard-thresholding the projection $\mathbf{W} \mathbf{x}{\left(m\right)}$ and keeping the $S_x$ entries of the largest magnitudes, setting the remaining low magnitude entries to zero. For this purpose, we define an intermediate vector ${\mathbf{f}}{\left(m\right)}  \triangleq \mathbf{W} \mathbf{x}{\left(m\right)} \in \mathbb{R}^L$ and denote by $\lambda_x$ the $S_x$-th largest magnitude among the set $\{ \vert f_1{\left(m\right)}  \vert , ..., \vert f_L{\left(m\right)}  \vert  \}$. Then the closed-form solution is achieved by applying to $\mathbf{f}{\left(m\right)} $ a hard-thresholding operator, which is defined as 
$ H_{\lambda_x} \! \left( \mathbf{f} \left(m\right)  \right) \! = \mathds{1}_{\vert f_l {\left(m\right)}  \vert \ge \lambda_x } {\mathbf{f}}{\left(m\right)}  ,   \forall m \in \left[ M \right], \forall l \in \left[ L\right] \!.$

\vspace{-2pt}

In \cite{Sohrab_WIFS2016}, the authors show that Sparse Ternary Coding (STC) outperformed classical binary hashing. Therefore, by considering the alphabet of sparse representation vectors as $\mathcal{A}  = \{ -1, 0, +1 \}$, we apply the ternary hash mapping to $H_{\lambda_x} \left( \mathbf{W} \mathbf{x} \left(m\right)  \right)$ as:
\vspace{-2pt}
\begin{equation}\label{Eq:OurEncoding}
\mathbf{a} \left(m\right) \triangleq T_{\lambda_x} \left(  \mathbf{W} \mathbf{x} \left(m\right)  \right) \in  {\{ -1, 0, +1 \}}^L , \;  \forall m \in \left[ M \right],
\vspace{-2pt}
\end{equation}
where 
$T_{\lambda_x} \left(  \mathbf{W} \mathbf{x} \left(m\right)  \right) =  \mathrm{sign}  \left( H_{\lambda_x} \left( \mathbf{W} \mathbf{x} \left(m\right) \right) \right) $. We denote, therefore, the space of public storage as ${\mathcal{A}}^{L \times M}$. For simplicity, through out this paper, we denote by $\mathcal{S} \subset \mathcal{X}^N$ the space of vectors in the signal (original) domain and by $\mathcal{T} \subset \mathcal{A}^L$ the space of vectors in the transform domain. Also, we denote by $\psi (. )$ the operator $T_{\lambda} \left(. \right)$ in general. 


Note that given the transform matrix $\mathbf{W}$ and sparse feature code $\mathbf{a}{(m)}$, the recovered feature vector is then simply $\mathbf{\widehat{x}}{\left(m\right)} = \mathbf{W}^\dagger \mathbf{a}{(m)}$. We are especially interested in square, i.e., $L = M$, and over-complete, i.e., $L > M$, transform matrices. In this paper, we just address the square sparsifying transform. 
Obviously, the projection $T_{\lambda_x} \left(  .  \right)$ introduces a certain loss of information. In order to measure the difference between the original data $\mathbf{x}{\left(m\right)}$ and the corresponding reconstructed data $\mathbf{\widehat{x}}{\left(m\right)}$ via encoded values ${\mathbf{a}}{\left(m\right)}$, we use the mean squared error (MSE) as our distortion measure. Also, since our points are random vectors, the distortion measure is given as
$D \left( \mathbf{x}{\left(m\right)}, \mathbf{\widehat{x}}{\left(m\right)} \right) =  \frac{1}{N}\,  \mathbb{E} \left[   {\| \mathbf{x}{\left(m\right)} -  \mathbf{\widehat{x}}{\left(m\right)} \|}_2 \right].$
Sparsity level $S_x$ should be chosen such to the specific distortion level. 


Our sparsifying transform learning is based on the classical Procrustes matrix problem \cite{schonemann1966generalized}. That is, we seek an orthonormal matrix $\mathbf{W} \in \mathbb{R}^{N \times N}$, which most closely transforms a given matrix $\mathbf{X} \in \mathbb{R}^{N \times M}$ into a sparse matrix $\mathbf{A} \in \mathbb{R}^{N \times M}$. Therefore, using the Frobenius norm, the problem is to find $\mathbf{W}$ minimizing ${ \| \mathbf{W}\mathbf{X} - \mathbf{A} \|}_F^2$ subject to $\mathbf{W} \mathbf{W}^T = \mathbf{I}$. For the square sparsifying transform, it can be shown that $\mathbf{W} = \mathbf{U} \mathbf{V}^T$, where the columns in $\mathbf{U}$ and $\mathbf{V}$ are left-hand and right-hand singular vectors for $\mathbf{A} \mathbf{X}^T$. 
Note that, in this case $\mathbf{W}^\dagger$ is simply $\mathbf{W}^T$. Our algorithm for the above minimization problem alternates between solving for $\mathbf{A} = \psi \left( \mathbf{W} \mathbf{X}\right)$ (sparse coding step) and $\mathbf{W} = \mathbf{U} \mathbf{V}^T$ (transform update step), whilst the other variables are kept fixed. 

In our framework, we set  the sparsifying transform matrix $\mathbf{W}$ to be provided, i.e., only authorized parties have access to it. However, even if $\mathbf{W}$ is compromised, our ambiguization mechanism prevents to reconstruct $\mathbf{x}$ from $\mathbf{a}$. 
In the general case, the matrix $\mathbf{W}$ is applied selectively to the data and shared only between the concerned parties. The plurality of different matrices $\left\{ \mathbf{W} \right\}$ is ensured by a key-base initialization of the sparsified Procrustes procedure, i.e., $\mathbf{W}^{\left(0\right)} = f \left( {\mathbf{K}} \right)$. We choose $f \left( {\mathbf{K}} \right) = \mathbf{U}_K$, where $\mathbf{U}_K$ is unitary matrix obtained form $\mathbf{U}_K \mathbf{\Sigma}_K \mathbf{V}_K^T = \mathbf{K}$.  
In general, given the database $\mathbf{X}  \in \mathbb{R}^{N \times M}$ and the shared key matrix $\mathbf{K} \in \mathbb{R}^{L \times N}$ as initial seed for generating $\mathbf{W}$, the \textit{public sparsifying transform matrix} $\mathbf{W} \in \mathbb{R}^{L \times N}$ can be easily learnt and accessed by authorized parties. Next, given $\lambda_x$ the encoded ambiguized data can be shared in public storage. Before our discussion on the imposed ambiguization for the server in Sec.~\ref{Sec:IV}, note that even if the server has the clean version of the transformed data, since it has no access to the shared key matrix $\mathbf{K}$, it cannot reconstruct the original data by the inverse projection. In Sec.~\ref{Sec:IV}, we will propose a simple method such that given the disclosed private key $\mathbf{K}$ and public sparsifying transform $\mathbf{W}$ at the server, it is infeasible to learn the structure of data.

\vspace{-8pt}

\subsection{Link to Quantized Embedding Methods}
\vspace{-2pt}

Many methods like \cite{boufounos2011secure} propose a similar privacy preserving framework based on:
\vspace{-5pt}
\begin{equation}\label{Eq:QuantizedEmbedd}
\mathbf{a} \left(m \right) = Q \left( \mathbf{W} \mathbf{x} \left(m\right) + \boldsymbol{\Delta} \right),  \forall m \in \left[ M \right],
\end{equation}
where $Q (. )$ stands for a scalar quantizer and $\boldsymbol{\Delta}$ is a dither that might be treated as a secret key and $\mathcal{A} = \left\{ -1, +1 \right\}$. There are several fundamental differences between \eqref{Eq:OurEncoding} and \eqref{Eq:QuantizedEmbedd}: \vspace{-2pt}
\begin{compactitem} 
\item[-]
Transform $\mathbf{W}$: In \eqref{Eq:QuantizedEmbedd}, $\mathbf{W}$ is a dimensionality reduction transform with random entries. In \eqref{Eq:OurEncoding}, the transform $\mathbf{W}$ is considered as an over-complete transform that extends the dimensions. Moreover, $\mathbf{W}$ in \eqref{Eq:OurEncoding} is trained using the described sparsifying Procrustes problem to ensure an optimal sparse representational that is information preserving in general, whereas $\mathbf{W}$ in \eqref{Eq:QuantizedEmbedd} might preserve the distances only under certain conditions of the Johnson-Lindenstrauss Lemma.
\item[-]
Codes $\mathbf{a}$: In \eqref{Eq:QuantizedEmbedd}, the codes are dense and binary, whereas in \eqref{Eq:OurEncoding} the codes are sparse and ternary, which form a basis of our ambiguization framework.
\end{compactitem}

\vspace{-8pt}

\subsection{Identification Stage}

\vspace{-3pt}

At the identification stage, given a query vector $\mathbf{y} \in \mathbb{R}^N$, sparsifying transform matrix $\mathbf{W}$ and sparsity level $S_y$, vector $\mathbf{b} \left(m\right) \triangleq  T_{\lambda_y} \left( \mathbf{W} \mathbf{y} \left(m\right) \right)\in  {\{ -1, 0, +1 \}}^L$ is calculated. We consider two hypothesis for our query, as follows:
\vspace{-2pt}
\begin{eqnarray}
\left\{ 
\begin{array}{ll}
\mathcal{H}_1: & \mathbf{y} = \mathbf{x} \left(m\right) + \mathbf{z} \, , \, \forall m \in \left[ M \right]\, ,   \\
\mathcal{H}_0: & \mathbf{y} = \mathbf{\acute{x}} \, ,
\end{array} \right.
\end{eqnarray}
where $\mathbf{\acute{x}} \in \mathbb{R}^N$ is a query vector out of the database. The block diagram of our model is depicted in Fig.~\ref{Fig:Model}. 

\vspace{-8pt}

\subsection{Practical Decoding}
\vspace{-2pt}

Our final decision is based on the $\gamma$-NN decoder, which seeks all $\{ \mathbf{a} \left(m\right) ,m \in \left[ M \right] \}$ NNs in the radius $\gamma L$ from the query $\mathbf{y}$ in order to produce a list of possible candidates as:
$\mathcal{L} \left( \mathbf{y} \right) = \left\{ m \in \left[M\right] : d_{\mathcal{A}} \left( \mathbf{a}\left(m \right) , \mathbf{b} \right) \leq \gamma L \right\},$
where $ d_{\mathcal{A}} \left( . , . \right)$ is our similarity measure in space $\mathcal{A}$.
We consider two decoding schemes, \textit{private} and \textit{public} decoding. In the private decoding, the client discloses his probe sparse representation, i.e., $T_{\lambda_y} \left( \mathbf{W} \mathbf{y} \left(m \right)\right)$, to the server completely. While in the public scheme, the client sends only positions of non-zero components in the sparse representation of his probe along with a fraction of ambiguous positions as $\mathbf{p}$. 

\vspace{-8pt}


\subsection{Isometric Mapping}
\vspace{-2pt}

In general, for a set $\mathcal{X}$, if we can introduce a metric (or distance function) $d_{\mathcal{X}}$ on $\mathcal{X} \times \mathcal{X}$ which obeying four fundamental axioms \cite{kreyszig1989introductory},
then a pair $\left( \mathcal{X}, d_{\mathcal{X}} \right)$ is said to be a metric space. In our framework we take the set of all ordered $N$-tuples of $\mathcal{S}$ as our original data, and the set of all ordered $L$-tuples of $\mathcal{T}$ as our transformed data. Let $\mathcal{S} = \left( \mathcal{S}, d_{\mathcal{S}}\right)$, $\mathcal{T} = \left( \mathcal{T}, d_{\mathcal{T}}\right)$ be metric spaces, then a mapping $\psi (.)$ of $\mathcal{S}$ into  $\mathcal{T}$ is said to be isometric, if $\psi(.)$ preserves distances.

\vspace{-2pt}

We are interested to preserve distances by our encoding scheme, i.e., there exists a sufficiently small $ \epsilon > 0$ such that for all $m \in \left[M\right]$, we have: 
\vspace{-4pt}
\begin{IEEEeqnarray}{lll}
\! \! \left(1- \epsilon \right)  d_{\mathcal{S}}  \left( \mathbf{x}\left(m\right) , \mathbf{y}\left(m\right) \right) \qquad \qquad \qquad \qquad \qquad\nonumber  \\
\qquad \qquad\quad \leq d_{\mathcal{T}} \left( \psi  \left( \mathbf{x}\left(m\right)  \right) ,  \psi  \left( \mathbf{y}\left(m\right)  \right) \right)  \leq  \nonumber \\
\quad \qquad \qquad  \qquad \qquad \qquad \left(1+ \epsilon \right)   d_{\mathcal{S}} \left( \mathbf{x} \left(m\right), \mathbf{y} \left(m\right) \right) \!.
 \vspace{-4pt}
 \end{IEEEeqnarray}
Assuming our metric is induced by the Euclidean norm, we have $d_{\mathcal{S}} \left(  \mathbf{x} ,  \mathbf{y} \right) =  {\| \mathbf{x} - \mathbf{y}  \|}_2, d_{\mathcal{T}} \left(  \mathbf{a} ,  \mathbf{b} \right) =  {\| \mathbf{a} - \mathbf{b}  \|}_2$.
Therefore, noting $ {\| \mathbf{x}{\left(m\right)} - \mathbf{y}{\left(m\right)} \|}_2^2 =   {\| \mathbf{z} \|}_2^2 \simeq N  \sigma^2_{\mathbf{z}}$ and considering a ternary sparse encoding, we have:
\vspace{-4pt}
\begin{equation}\label{dist_pres_tsc}
 \left(1- \epsilon \right) {\| \mathbf{z} \|}_2^2 \leq {\| \mathbf{a}{\left(m\right)} - \mathbf{b}\left(m\right) \|}_2^2 \leq  \left(1 + \epsilon \right) {\| \mathbf{z} \|}_2^2.
 \vspace{-5pt}
\end{equation}

\begin{figure}[t!]
\centering
\includegraphics[width=5.65cm, height=2.7cm]{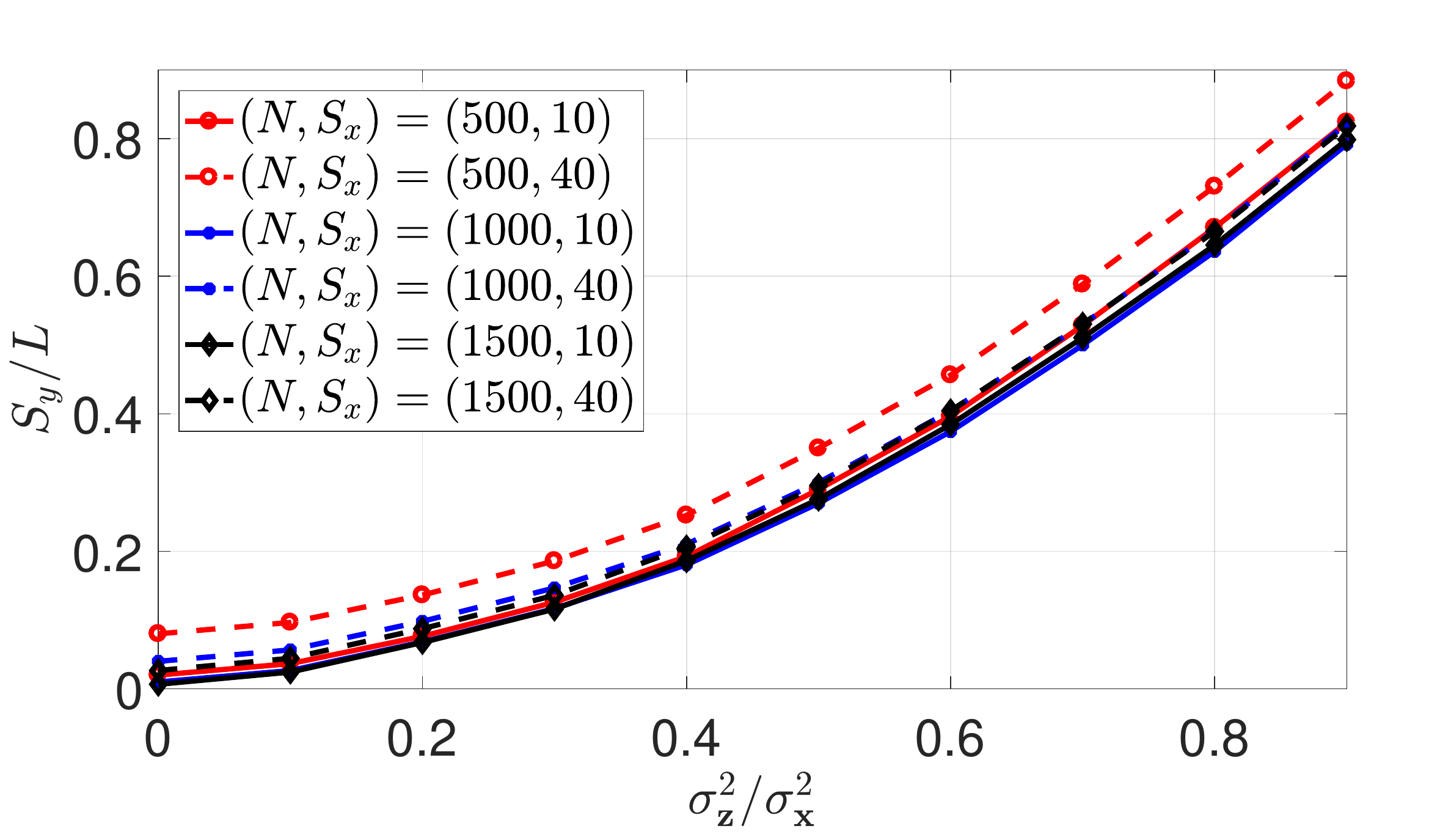}
\vspace{-3pt}
\caption{The relation between the required sparsity level $S_y$ normalized to code length $L\!  = \! N$ versus the variance of noise $\sigma^2_{\mathbf{z}}$ normalized to $\sigma^2_{\mathbf{x}}$ in order to preserve distances in the ternary sparse domain, for $M \!  \! =  \! 20000$ i.i.d random vectors with mean zero and variance $\sigma^2_{\mathbf{x}}$, with three different dimension of feature vectors $N  \! =  \! 500, 1000, 1500$ and two sparsity level $S_x  \! = \! 10, 40$. }
\vspace{-16pt}
 \label{Fig:ReqSy}
\end{figure}

\vspace{-3pt}
Before discussion on the above inequality, we consider lower and upper bounds of distances, which we will use in Section \ref{Sec:IV}.
In the original signal domain, considering query vector $\mathbf{y}$ in the general case, the lower and upper bounds of distances between the vector $\mathbf{x} \left(m\right),  m \in \left[ M \right]$ in the database and query $\mathbf{y}$ can be given as
$N  {\sigma}^{2}_{\mathbf{z} }  \simeq {\| \mathbf{z} \|}_2^2  \leq  {\| \mathbf{x}{\left(m\right)} - \mathbf{y} \|}_2^2 
 \leq  {\| \mathbf{x}{\left(m\right)} \|}_2^2 + {\|  \mathbf{y} \|}_2 ^2 \simeq N \! \left( \sigma^2_{\mathbf{x} }  +    \sigma^2_{\mathbf{y}}  \right)\!$.

\vspace{-3pt}

In the transform domain with the ternary encoding, we have 
$ \mathbf{W} \mathbf{x}{\left(m\right)}\!   =  \!\! T_{\lambda_x} \left( \mathbf{W} \mathbf{x}{\left(m\right)}  \right) \! +\! \mathbf{e}_{\mathbf{x}}= {\mathbf{a}}{\left(m\right)} + \mathbf{e}_{\mathbf{x}} $ and 
$ \mathbf{W} \mathbf{y}{\left(m\right)}\!    =  \!\! T_{\lambda_y} \left( \mathbf{W} \mathbf{y}{\left(m\right)}  \right) \! +\!  \mathbf{e}_{\mathbf{y}} ={\mathbf{b}}{\left(m\right)}  + \mathbf{e}_{\mathbf{y}}$,
where $\mathbf{e}_{\mathbf{x}}$ and $\mathbf{e}_{\mathbf{y}}$ are approximation errors of the ternary hash mapping corresponding to $\mathbf{x}{\left(m\right)}$ and $\mathbf{y}{\left(m\right)}$, respectively. Therefore, the upper bound on ${\| \mathbf{a} \left(m\right) \! \! - \!   {\mathbf{b}}{\left(m\right)}  \|}_2$ is given as:
\vspace{-3pt}
\begin{eqnarray} 
{\| \mathbf{a}\! \left(m\right) \! \! - \! \!  {\mathbf{b}}{\left(m\right)}  \|}_2^2 \leq \! {\| \mathbf{W} \|}_{\!F}^2  {\| \mathbf{x}{\left(m\right)} \! - \! \mathbf{y}{\left(m\right)} \|}_2^2  \! + \! { \|  \mathbf{e}_{\mathbf{x}}  \! -\!  \mathbf{e}_{\mathbf{y}}   \|}_2^2.
\end{eqnarray}
However, the above upper bound is not tight, since it does not consider properties of vectors in the transform domain. Also, it is based on the error terms of approximation, which may not give a clear sense of the upper bound at first glance.   
Considering the sparse ternary property of vectors, we have ${\|\mathbf{a} \left(m\right)  - \mathbf{b} \left(m\right)  \|}_2^2 = {\| \mathbf{a} \left(m\right)  \|}_2^2 + {\|  \mathbf{b} \left(m\right)  \|}_2^2 - 2 \langle \mathbf{a} \left(m\right)  , \mathbf{b} \left(m\right)  \rangle$ and $\vert  \langle \mathbf{a} \left(m\right)  , \mathbf{b} \left(m\right) \rangle  \vert \leq \min \{ S_x, S_y \}$, the lower and upper bounds on distance are given as:
\vspace{-5pt}
\begin{IEEEeqnarray}{lll}
S_x + S_y - 2 \min \{ S_x, S_y \} \qquad \qquad \qquad \qquad \qquad \qquad \nonumber  \\
\qquad \qquad\qquad \leq  {\| \mathbf{a} \left(m\right)  - \mathbf{b} \left(m\right)  \|}_2^2  \leq  \nonumber \\
\qquad \qquad \qquad \qquad \qquad  \quad S_x + S_y + 2 \min \{ S_x, S_y \}.
\vspace{-5pt}
\end{IEEEeqnarray}
This means that the upper bound on distances in the transform domain are restricted by sparsity levels.

\vspace{-4pt}
In order to preserve distances, given the variance of the measurement noise $\sigma_{\mathbf{z}}^2$ and sparsity level $S_x$\footnote{Sparsity level $S_x$ obtained based on the desired distortion level imposed by sparse approximation.}, we have to set $\left(1- \epsilon \right) \left( N \sigma_{\mathbf{z}}^2 \right) \leq S_x + S_y - 2 \langle \mathbf{a} \left(m\right)  , \, \mathbf{b} \left(m\right)  \rangle   \leq  \left(1 + \epsilon \right) \left( N \sigma_{\mathbf{z}}^2\right)$. So,  the smallest integer sparsity level for approximation of noisy query $\mathbf{y} \left(m\right)$ is given as $S_y = \lfloor\,   N \sigma_{\mathbf{z}}^2 + 2 \langle \mathbf{a} \left(m\right)  , \, \mathbf{b} \left(m\right)  \rangle - S_x \, \rfloor$. 
Note that if $\mathbf{z} = \mathbf{0}$, then $S_x = S_y$ since in this condition $\mathbf{a} \left(m\right)  =\mathbf{b} \left(m\right) $, therefore $ \langle \mathbf{a} \left(m\right)  , \mathbf{b} \left(m\right)  \rangle = S_x$. However, the above equation is implicit since $\mathbf{b} \left(m\right) $ depends on the sparsity level $S_y$, and vice versa. But, we can obtain it recursively. Fig.~\ref{Fig:ReqSy} depicts the required sparsity level $S_y$ in order to preserve distances in the signal and transform domains. 
Due to the lack of space, we cannot provide our discussion about thresholding functions $H_{\lambda_x}$ and $H_{\lambda_y}$ as well as our analytic approach for distortion analysis. The more detailed analytic results will be presented in an extended version of this paper. 

\vspace{-7pt}


\section{Ambiguization Analysis}\label{Sec:IV}
\vspace{-3pt}

\begin{figure}[!t] 
    \centering%
    \begin{subfigure}[b]{0.23\textwidth}
        \includegraphics[width=3.85cm, height=1.76cm]{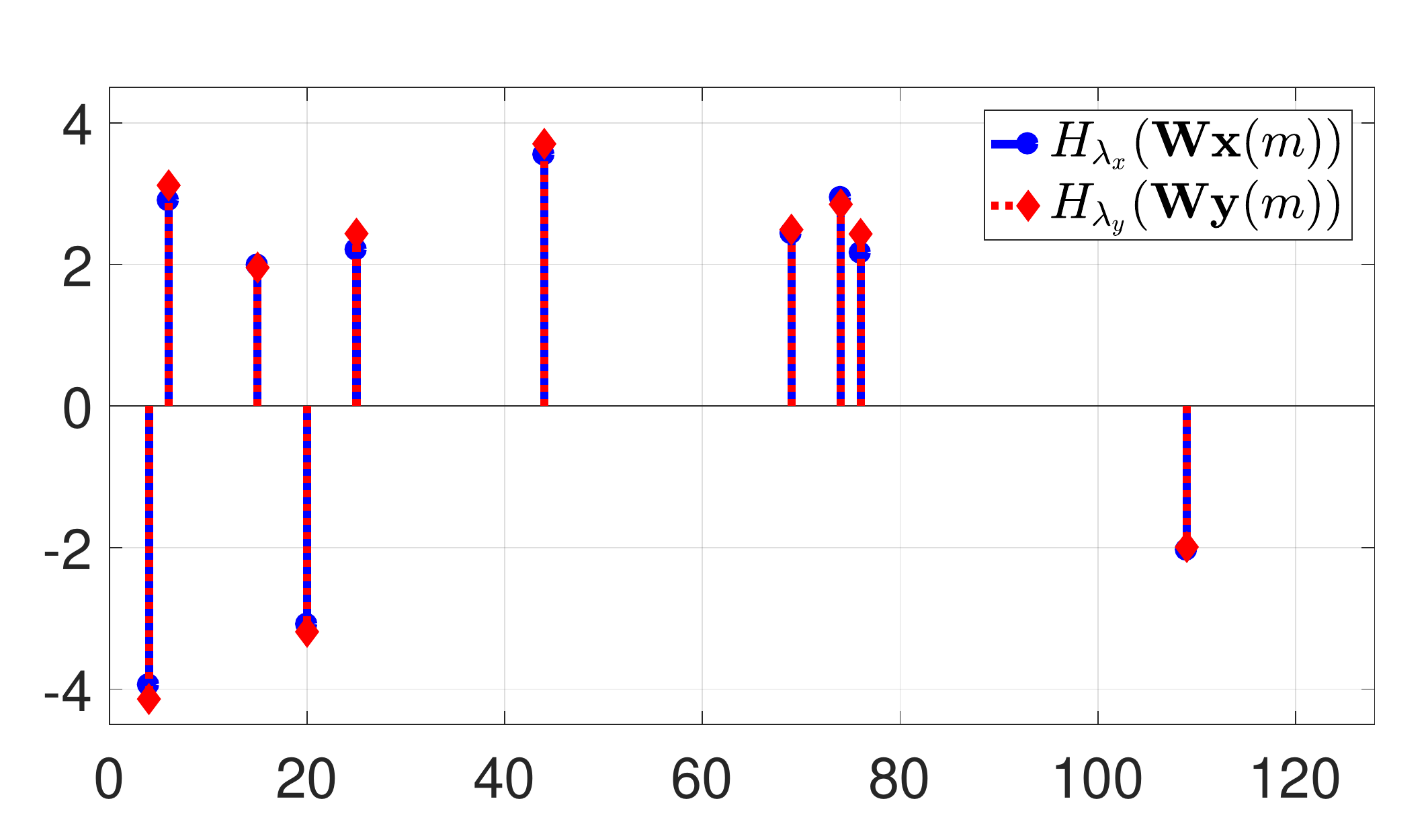}
          \vspace{-11pt}
        \caption{ }
        \label{fig:MesurementNoiseSprRepa}
    \end{subfigure}
    ~ 
    \begin{subfigure}[b]{0.23\textwidth}
        \includegraphics[width=3.85cm, height=1.76cm]{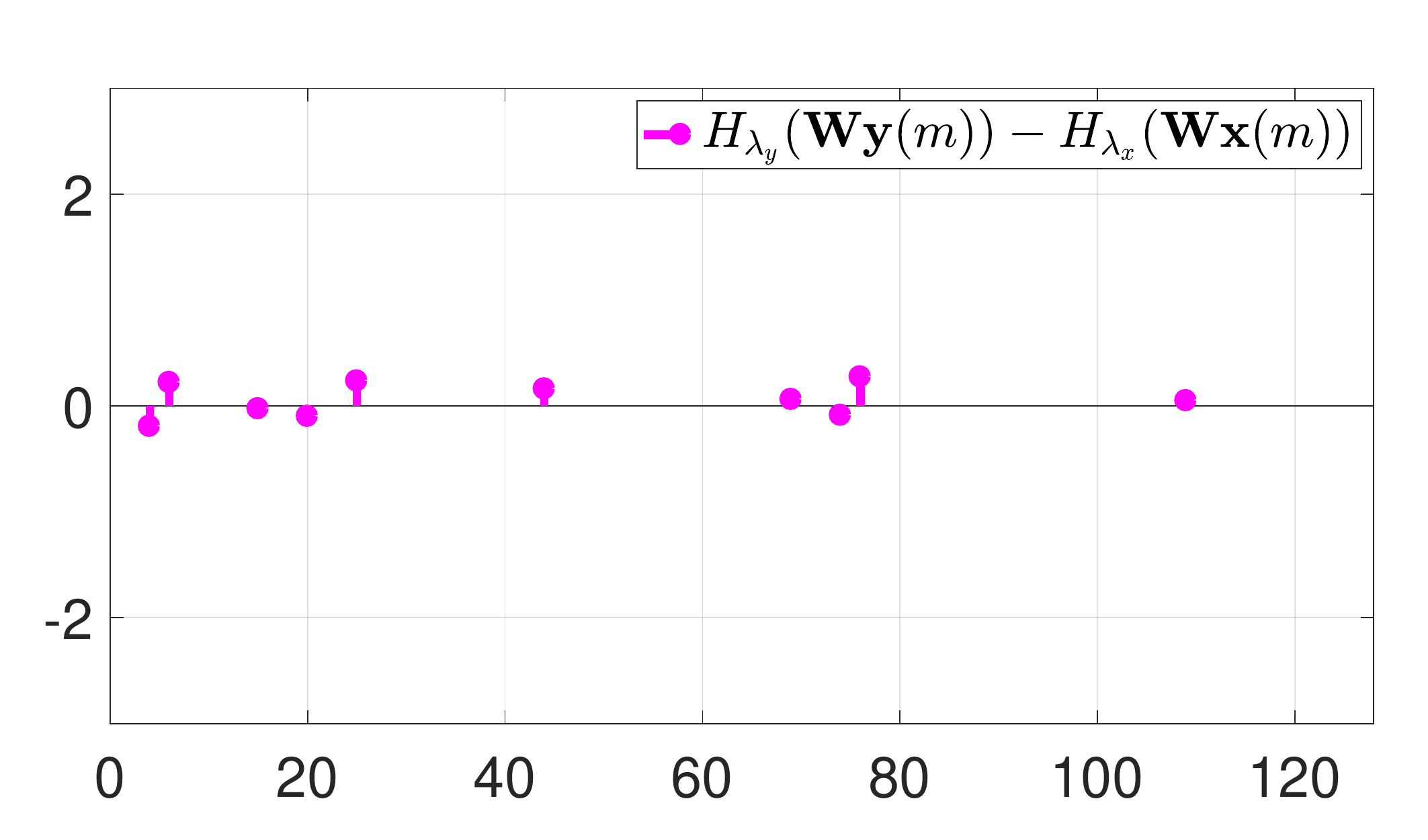}
          \vspace{-11pt}
        \caption{ }
        \label{fig:MesurementNoiseSprRepb}
    \end{subfigure}
    
     \vspace{-10pt}
    \begin{subfigure}[b]{0.23\textwidth}
        \includegraphics[width=3.85cm, height=1.76cm]{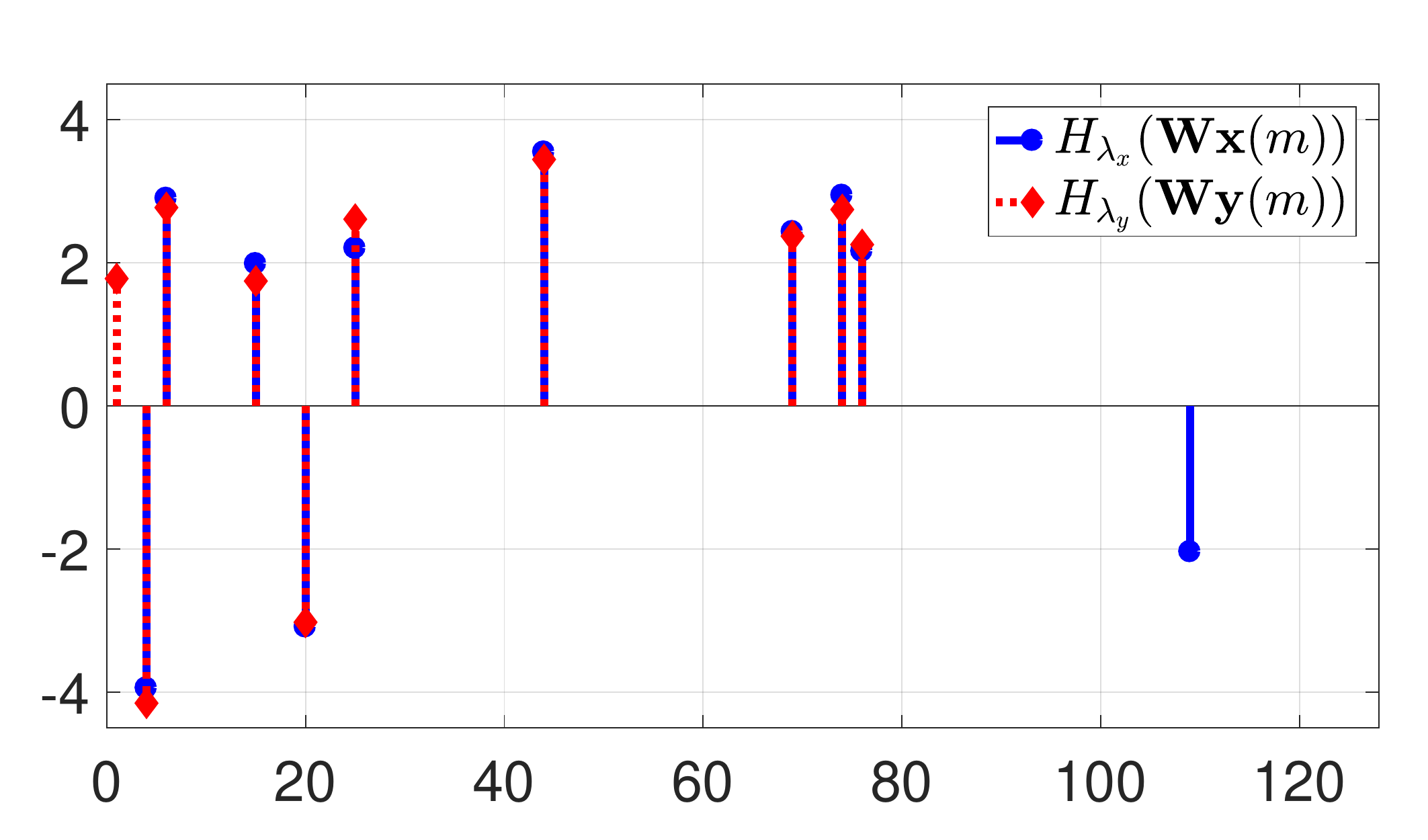}
          \vspace{-11pt}
         \caption{}
        \label{fig:MesurementNoiseSprRepc}
    \end{subfigure}
    ~ 
    \begin{subfigure}[b]{0.23\textwidth}
        \includegraphics[width=3.85cm, height=1.76cm]{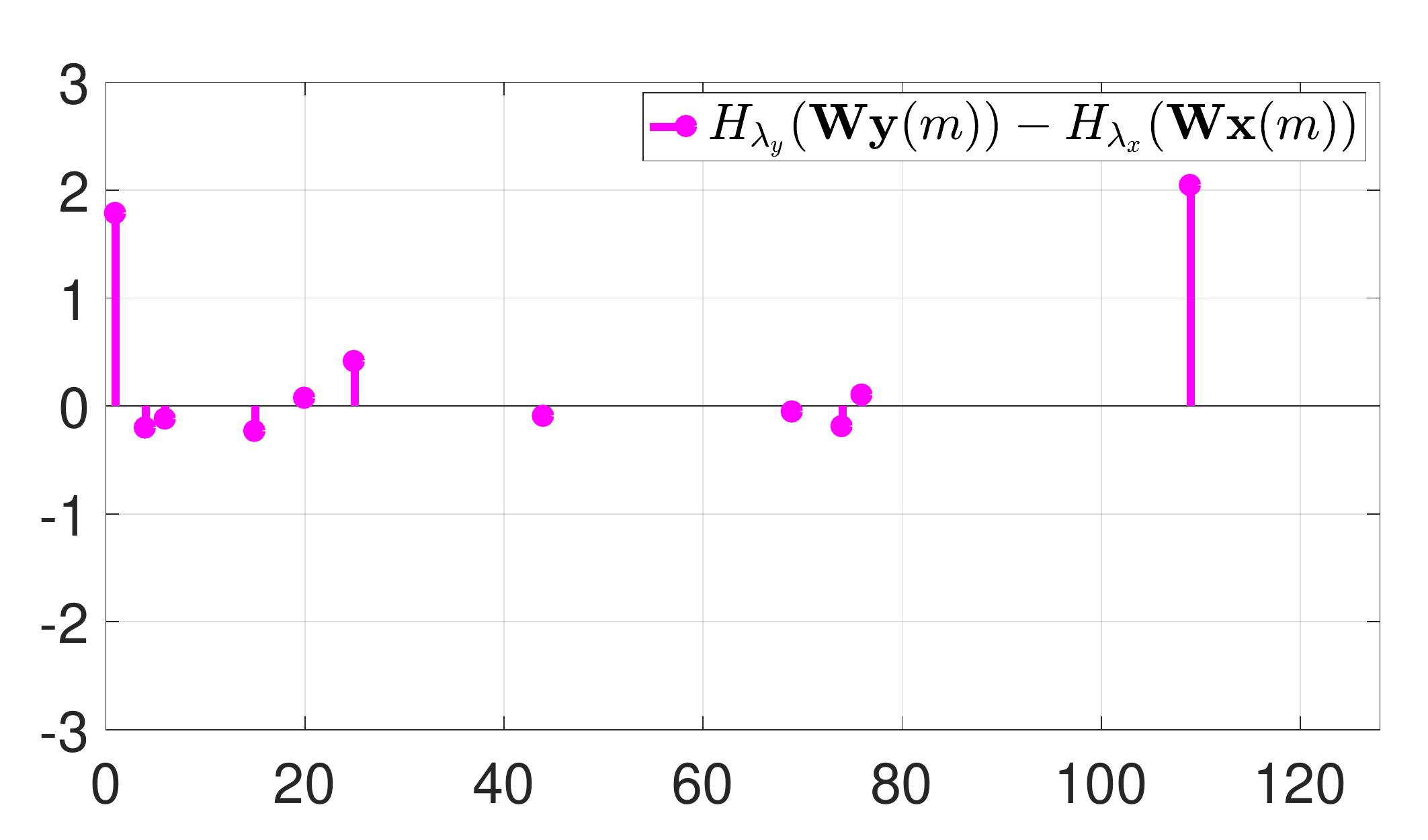}
          \vspace{-11pt}
        \caption{}
        \label{fig:MesurementNoiseSprRepd}
    \end{subfigure}
    
    \vspace{-10pt}
        \begin{subfigure}[b]{0.23\textwidth}
        \includegraphics[width=3.85cm, height=1.76cm]{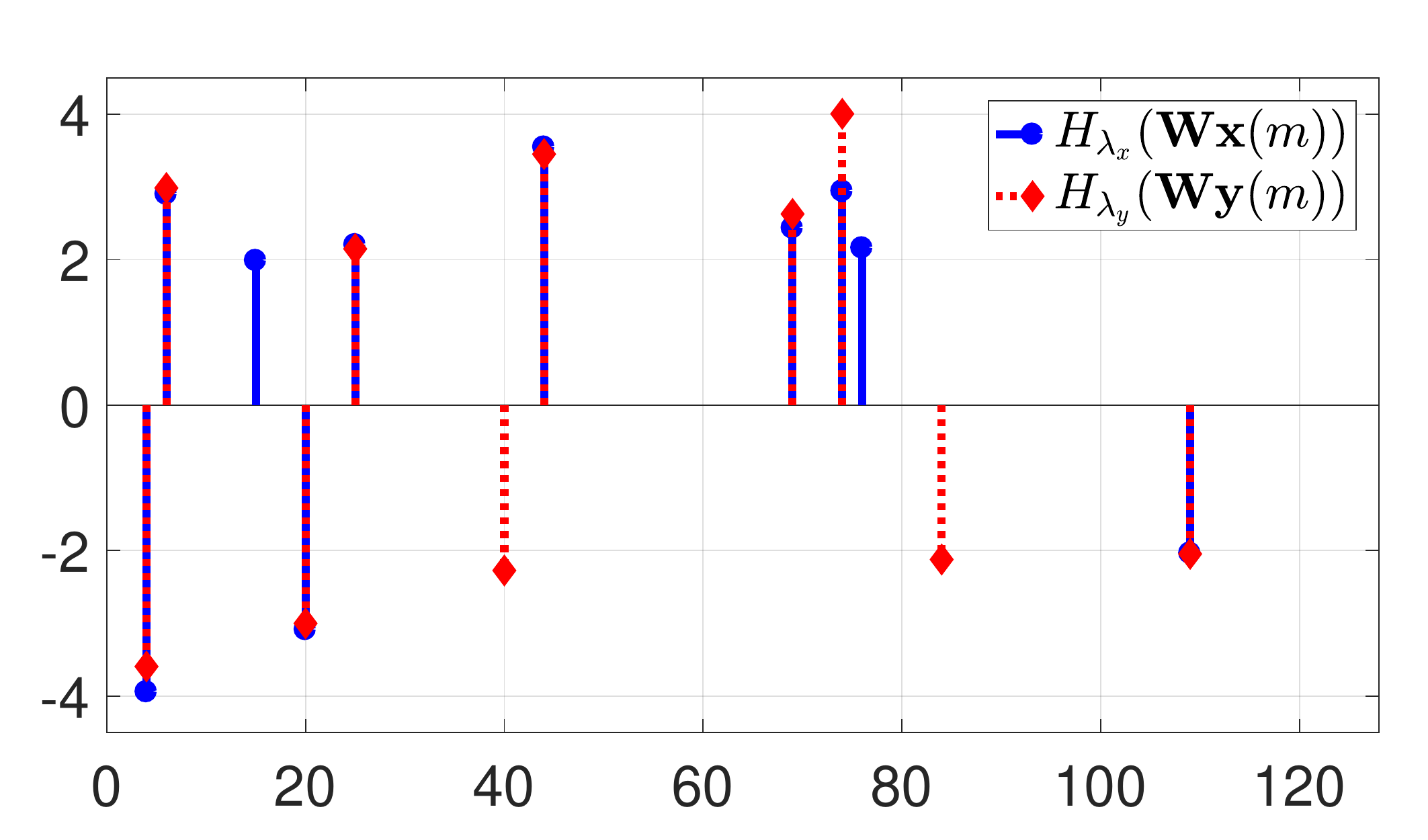}
          \vspace{-11pt}
        \caption{}
        \vspace{-13pt}
        \label{fig:MesurementNoiseSprRepe}
    \end{subfigure}
    ~ 
    \begin{subfigure}[b]{0.23\textwidth}
        \includegraphics[width=3.85cm, height=1.76cm]{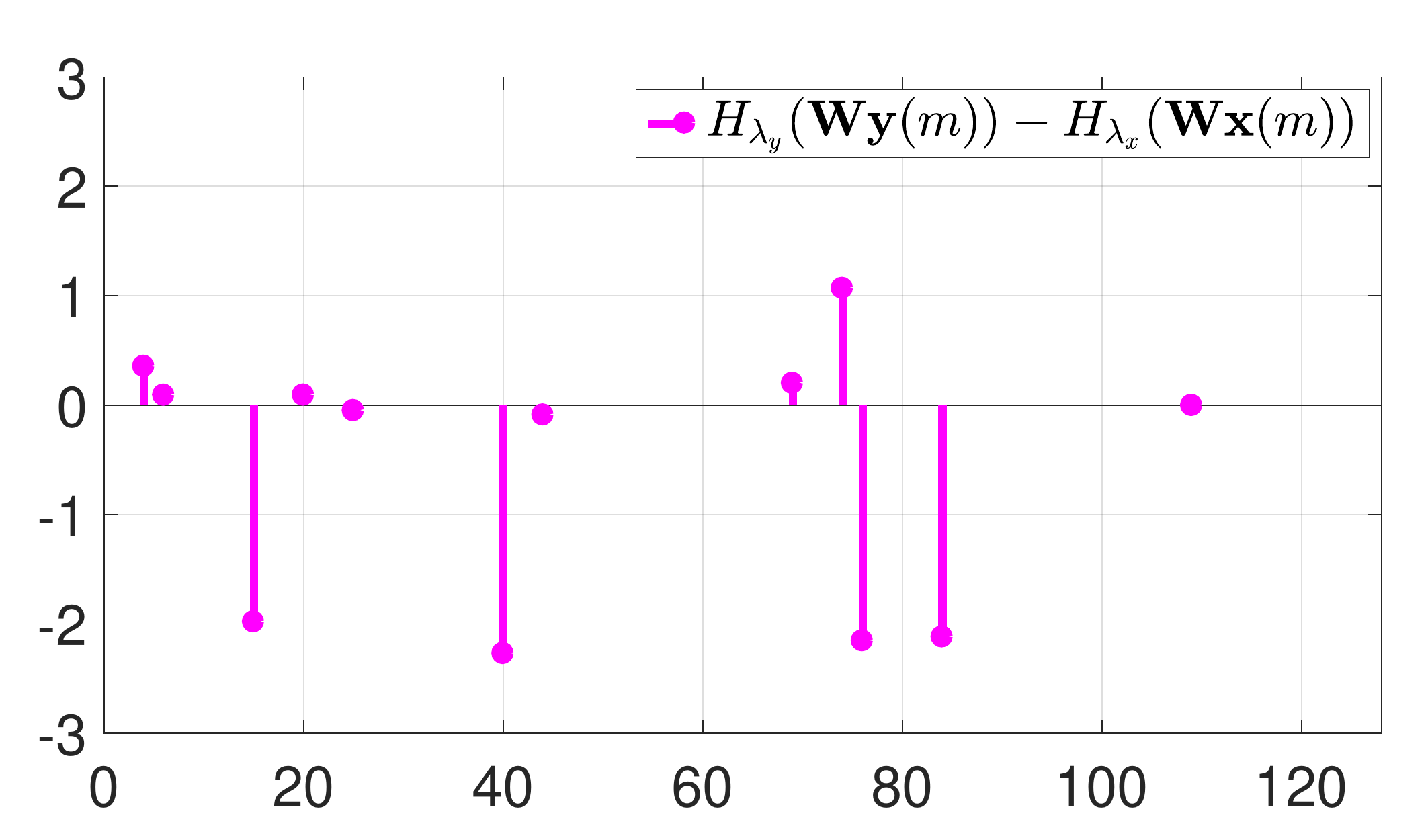}
      \vspace{-11pt}
        \caption{}
        \vspace{-13pt}
        \label{fig:MesurementNoiseSprRepf}
    \end{subfigure}
    \caption{Impact of the variance of measurement noise on the components of $H_{\lambda_x}  \! (\mathbf{W}\mathbf{x} (m ))$ and corresponding $H_{\lambda_y} \!  (\mathbf{W}\mathbf{y} (m ))$, for randomly selected $m \! \!   \in  \!  \! \left[ 500 \right]$, with $S_x \!   \! =  \! S_y  \! = \! 10$, $N   \! \! = \!  \!  L   \! \! =  \!  \!  128$, $\sigma^2_{\mathbf{x}}  \!  \! = \!  \!  1$.  Fig.~(\ref{fig:MesurementNoiseSprRepa}), Fig.~(\ref{fig:MesurementNoiseSprRepc}) and Fig.~(\ref{fig:MesurementNoiseSprRepe}) represent the components of $H_{\lambda_x} \!  (\mathbf{W}\mathbf{x}\left(m\right))$ and $H_{\lambda_y} \!  (\mathbf{W}\mathbf{y}\left(m\right))$ with $\sigma^2_{\mathbf{z}}  \!  \! =  \! \!  0.15$, $\sigma^2_{\mathbf{z}}  \!  \! =  \! \!  0.2$ and $\sigma^2_{\mathbf{z}}  \!  \! =  \!  \! 0.4$, respectively. Fig.~(\ref{fig:MesurementNoiseSprRepb}), Fig.~(\ref{fig:MesurementNoiseSprRepd}) and Fig.~(\ref{fig:MesurementNoiseSprRepf}) indicate the corresponding errors.}
    \vspace{-16pt}
 \label{Fig:MesurementNoiseSprRep}
\end{figure}

In this section, we provide our results for ambiguization of public storage. In order to make our ambiguization approach more clear, first we discuss our model without ambiguization noise. In Fig.~\ref{Fig:MesurementNoiseSprRep}, we illustrates impact of measurement noise on the sparse approximation with a hard-thresholding operator for small sparsity levels $S_x = S_y = 10$~\footnote{The same sparsity levels are not optimal to preserve distances. We just intend to make sense of the influence of measurement noise on clean data.}. We see that small measurement noise just changes the magnitude of the components, while the indices of sparse components gradually miss by increasing the variance of the noise.

\vspace{-7pt}

\subsection{Ambiguization in Sparsified Transform Domain}

\vspace{-4pt}

In this section, we propose a simple scheme for imposing ambiguization at the server. In previous sections, we use $\mathcal{T}$ as space of $L$-dimensional vectors in the transform domain. Let $\mathcal{U}, \mathcal{V} \subseteq \mathcal{T}$ be two subspaces such that $\mathcal{T} = \mathcal{U} + \mathcal{V}$. So, every vector $\mathbf{a} \in \mathcal{T}$ has at least one expression as $\mathbf{a} = \mathbf{u} + \mathbf{v}, \mathbf{u} \in \mathcal{U}, \mathbf{v} \in \mathcal{V}$. If we have $\mathcal{U} \cap \mathcal{V} = \left\{ \mathbf{0}  \right\}$, then every vector $\mathbf{a} \in \mathcal{T}$ has the \textit{unique} expression $\mathbf{a} = \mathbf{u} + \mathbf{v}, \mathbf{u} \in \mathcal{U}, \mathbf{v} \in \mathcal{V}$ and we write $\mathcal{T} = \mathcal{U} \oplus \mathcal{V}$. Also, $ \mathcal{T}$ is called the direct sum of $\mathcal{U}$ and $\mathcal{V}$. 
Now, let  $\mathcal{U}$ be the space of non-zero components of $\mathcal{T}$ and $\mathcal{V}$ be the space of zero components of $\mathcal{T}$. 
The idea of our ambiguization method is to set ambiguization noise $\mathbf{n}$ such that $\mathbf{n} \in  \mathcal{V}$. 
Furthermore, since $\left( \mathcal{T}, \langle ., . \rangle \right)$ is an inner product space and $\mathcal{V} = \mathcal{U}^\bot \triangleq \left\{ v \in \mathcal{V} : \langle u, v \rangle = 0, \forall u \in \mathcal{U} \right\}$, $\mathcal{T} = \mathcal{U} \oplus \mathcal{U}^\bot$ is orthogonal direct sum of $ \mathcal{U}$ and $ \mathcal{U}^\bot$. It is clear that $\mathrm{dim}~\mathcal{T} = L$, $\mathrm{dim}~\mathcal{U} = S_x$ and $\mathrm{dim}~\mathcal{U}^\bot = L - S_x$. We denote by $S_{n_s}$ the sparsity level of ambiguization noise at the server. Note that $0 \leq S_{n_s} \leq L - S_x$. 

\vspace{-4pt}

Intuitively, when the number of choices increases, the server uncertainty increases, and so does the entropy measure proportionally to {\footnotesize $\log_2 \displaystyle \binom {S_x + S_{n_s}} {S_x}$}. By adding ambiguization noise $\mathbf{n}$ to the zero components of the sparse representation of the clean data, we increase uncertainty at the server side. 
Clearly, we have a trade-off between utility and privacy. By setting sparsity levels such that $S_x \rightarrow 0$ and $S_{n_s} \rightarrow L$ we reach maximal privacy and loose our utility in the sense of a stable reconstruction, correct decision and also sparse representation. On the other hand, by choosing $S_x \rightarrow L$ and $S_{n_s} \rightarrow 0$ we reach maximal utility, however, we loose our privacy. However, it should be noted that authorized client shares a common secrecy with the data owner, i.e., he knows a set of secret positions in contrast to the server. Therefore, the addition of extra false components in the orthogonal positions does not harm the client as much as the server.

\begin{figure}[t]
    \centering
     \hspace{-8pt}
        \begin{subfigure}[h]{0.24\textwidth}
        \includegraphics[width=4.75cm, height=3.61cm]{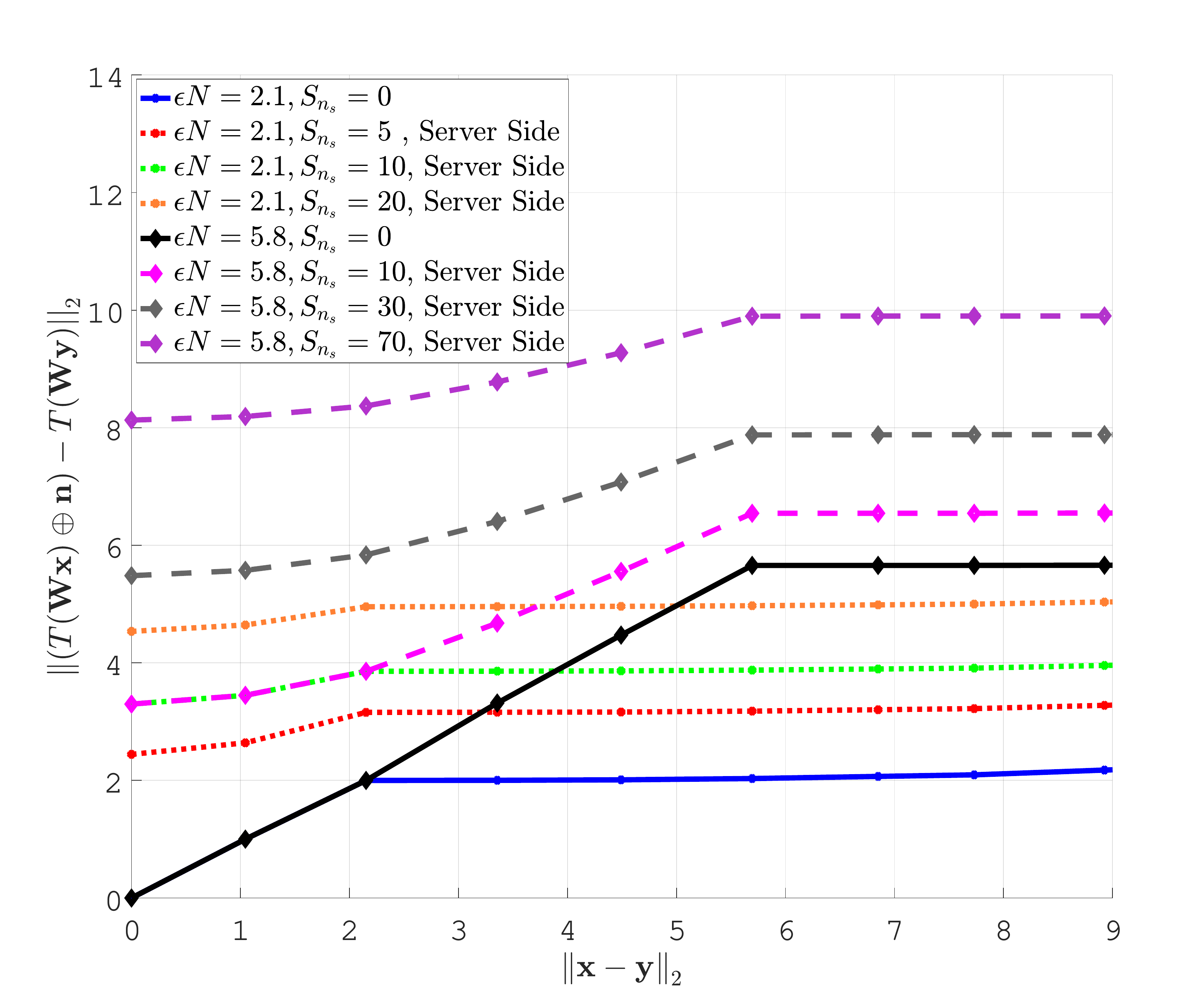}%
           \vspace{-5pt}
        \caption{Server Side}
           \vspace{-10pt}
        \label{fig:DisPreServerSide}
    \end{subfigure}%
 ~
       \begin{subfigure}[h]{0.24\textwidth}
        \includegraphics[width=4.75cm, height=3.611cm]{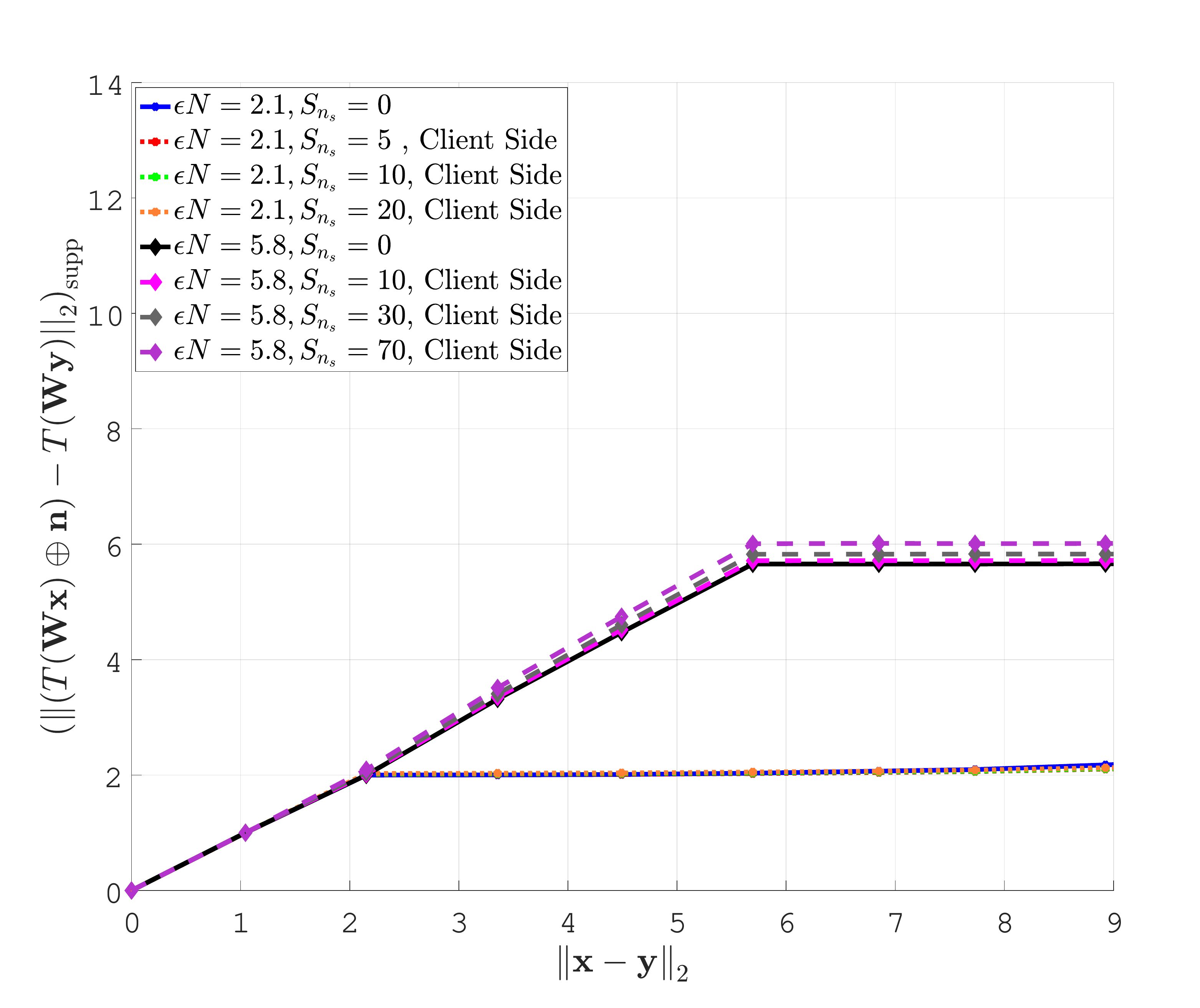}%
        \vspace{-5pt}
        \caption{Client Side}
           \vspace{-10pt}
        \label{fig:DisPreQuerySide}
    \end{subfigure}
    \caption{Impact of ambiguization at (a) Server side and (b) Client side, for two different desired radius of ${\| \mathbf{x} - \mathbf{y} \|}_2 = \epsilon N = 2.1, 5.8$ in the original domain with different ambiguization noise sparsity levels $S_{n_s}$.}
    \vspace{-15pt}
    \label{Fig:DisPre}
\end{figure}

\vspace{-2pt}

Fig.~\ref{Fig:DisPre} depicts the effect of ambiguization on the distances at the server side and client side. Considering no ambiguization noise on the sparse representation of public data, the solid "blue" and "black" lines illustrate our desired preserved distances corresponding to $\| \mathbf{x} - \mathbf{y} \| = 2.1$ and $\| \mathbf{x} - \mathbf{y} \| = 5.8$, respectively. Hence, on the client side, we cannot recover any information about the NNs hash signals that are far apart, since our approach only preserves distances up to a radius, as determined by the sparsity level of the approximation.
As it is shown, by imposing ambiguization noise, all distances from the server viewpoint go to constant value, i.e., all vectors in the public domain seem equally from the server standpoint. However, from the client side we can effectively preserve distances up to the designed radius, just by computing distances on non-zero components of the sparse representation of the query, i.e., on $\mathrm{supp} \left( T_{\lambda_y} \left( \mathbf{W} \mathbf{y}(m)  \right) \right)$. Note that, regarding to our discussion in Sec.~\ref{Sec:III}, the upper bound on distances in the transform domain is restricted by sparsity levels.

\begin{figure}[t]
    \centering
        \begin{subfigure}[h]{0.24\textwidth}
        \includegraphics[scale=0.165]{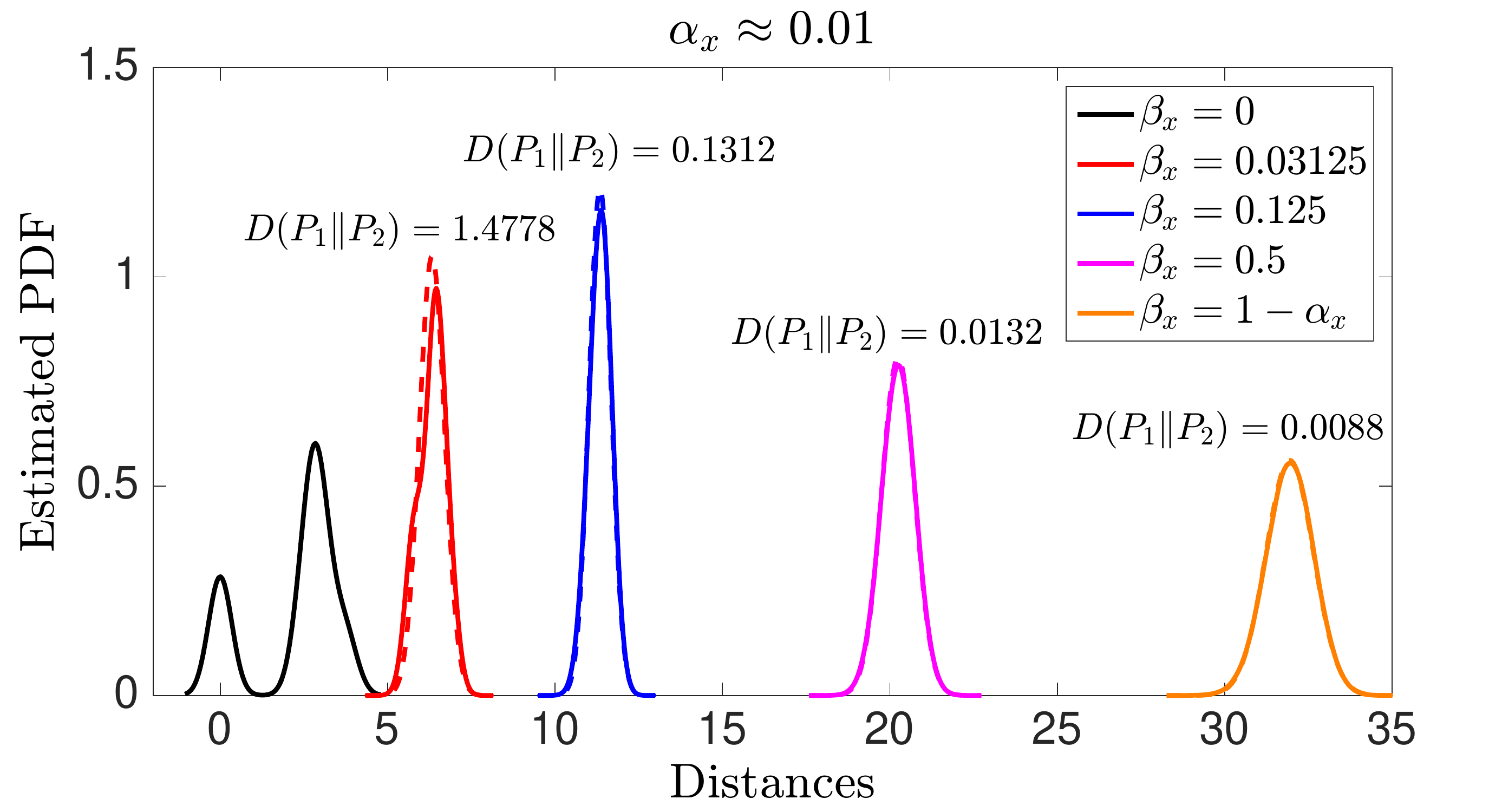}%
           \vspace{-6pt}
        \caption{ }
           \vspace{-7pt}
        \label{fig:pdf1}
    \end{subfigure}%
 ~
       \begin{subfigure}[h]{0.24\textwidth}
        \includegraphics[scale=0.165]{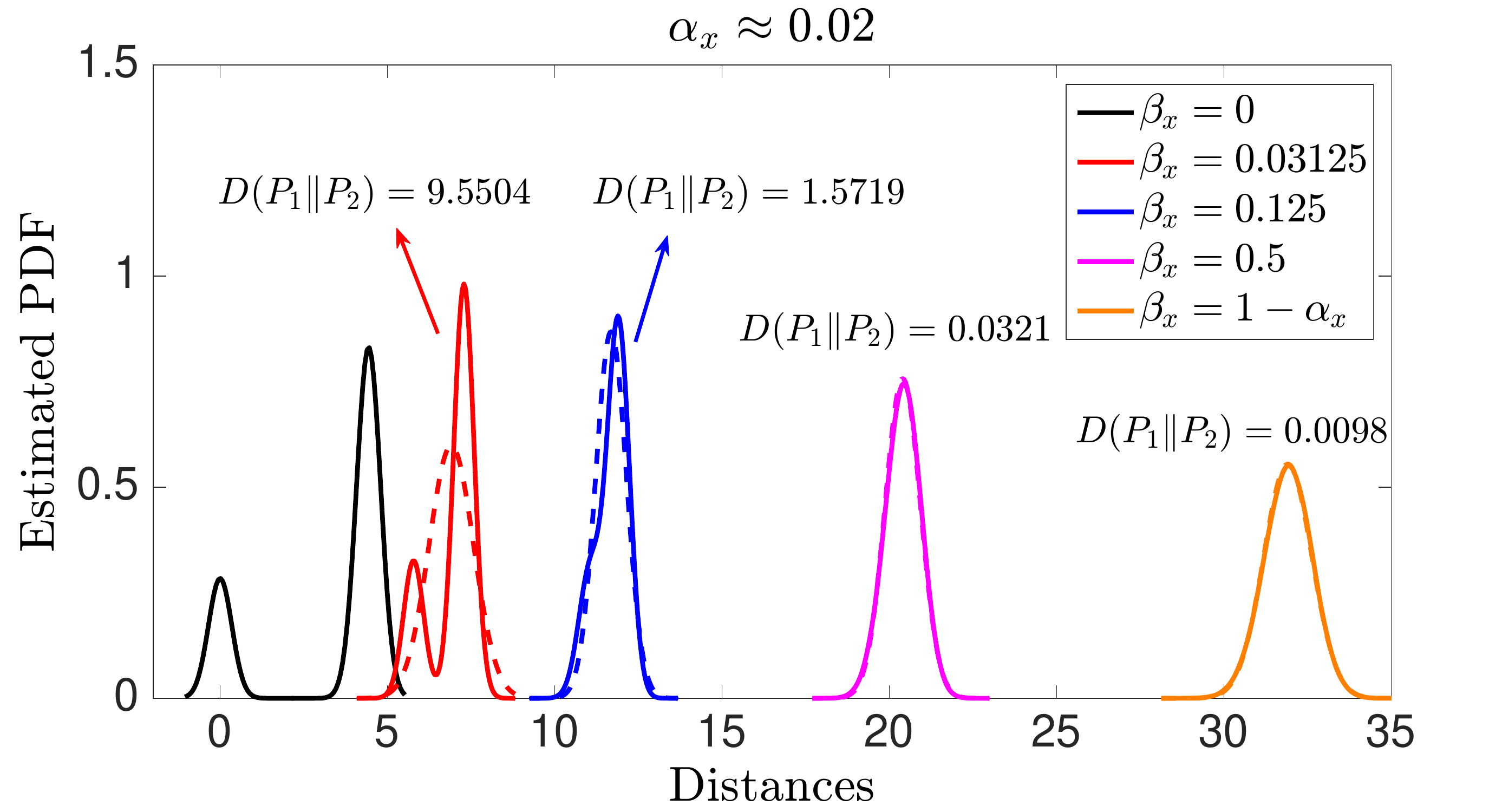}%
        \vspace{-6pt}
        \caption{ }
           \vspace{-7pt}
        \label{fig:pdf2}
    \end{subfigure}
 
            \begin{subfigure}[h]{0.24\textwidth}
        \includegraphics[scale=0.165]{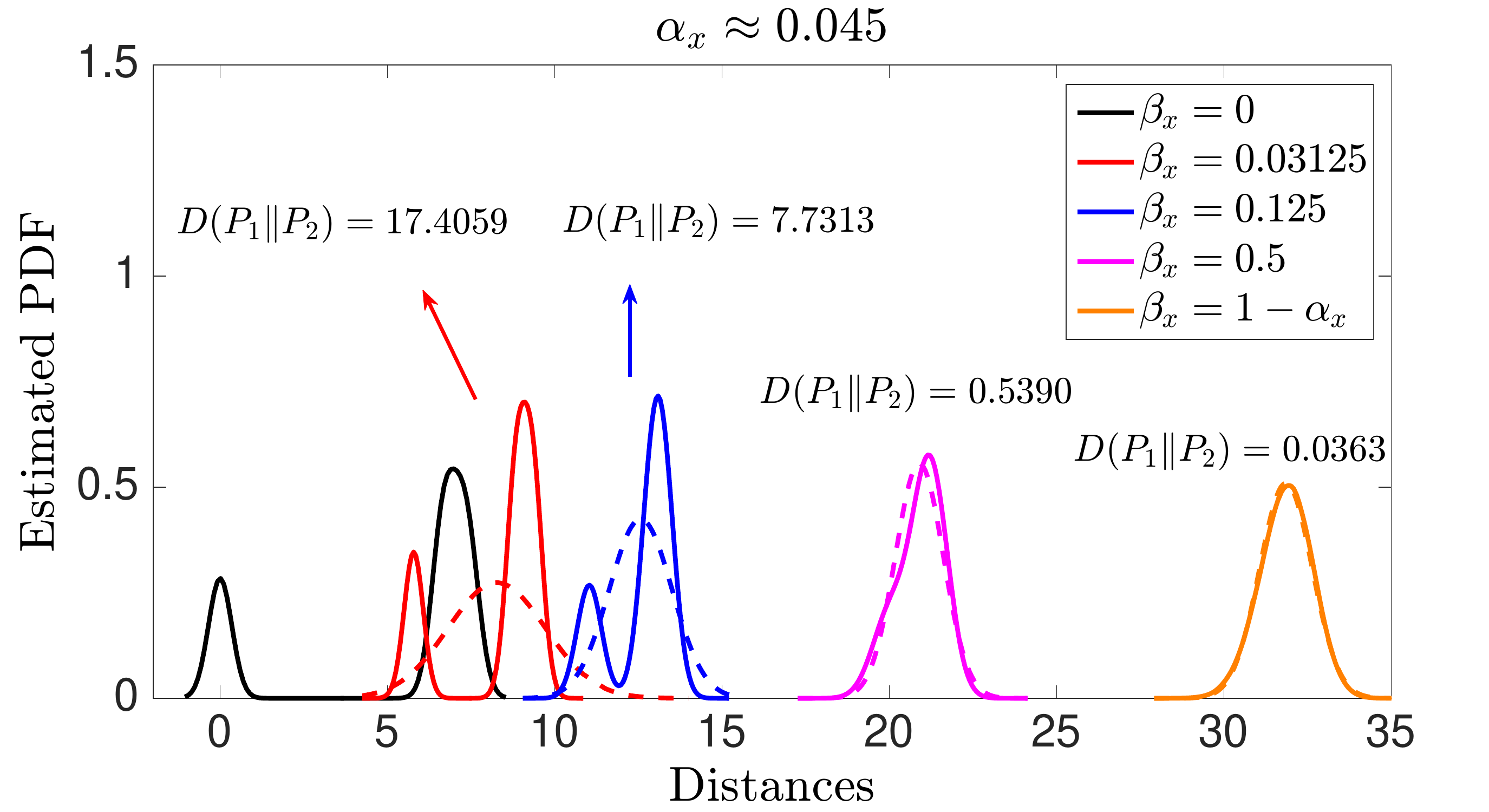}%
           \vspace{-6pt}
        \caption{ }
           \vspace{-13pt}
        \label{fig:pdf3}
    \end{subfigure}%
 ~
       \begin{subfigure}[h]{0.24\textwidth}
       \includegraphics[scale=0.165]{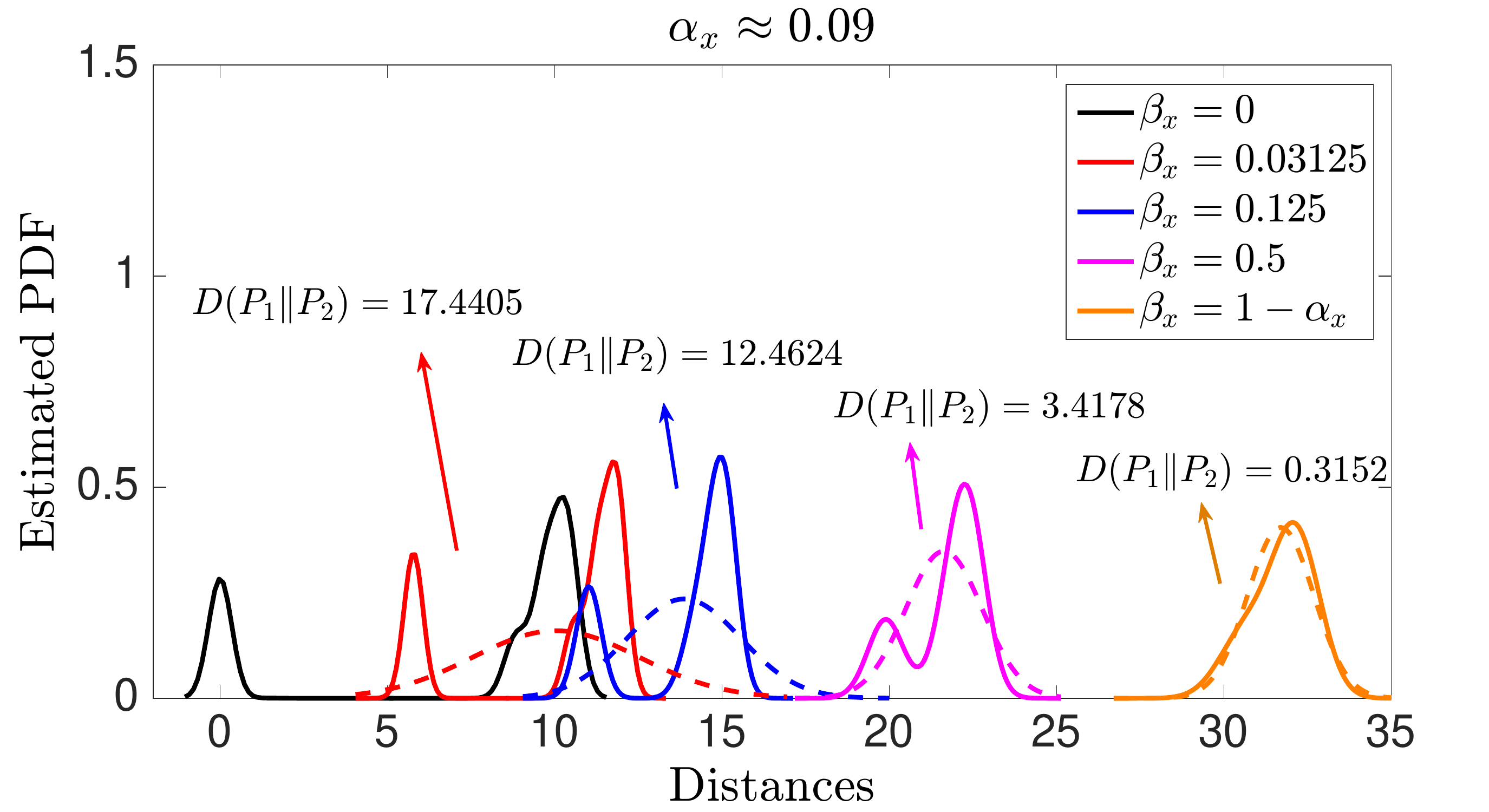}%
        \vspace{-6pt}
        \caption{ }
           \vspace{-13pt}
        \label{fig:pdf4}
    \end{subfigure}
    \caption{Estimated PDFs of pairwise distances in transform domain with different sparsity ratios and privacy leak estimates $D(P_1 \| P_2)$ for the corresponding cases.}
    \vspace{-20pt}
    \label{Fig:PDF}
\end{figure}

\vspace{-3pt}

\begin{figure}[!t]
\centering
\includegraphics[scale=0.4]{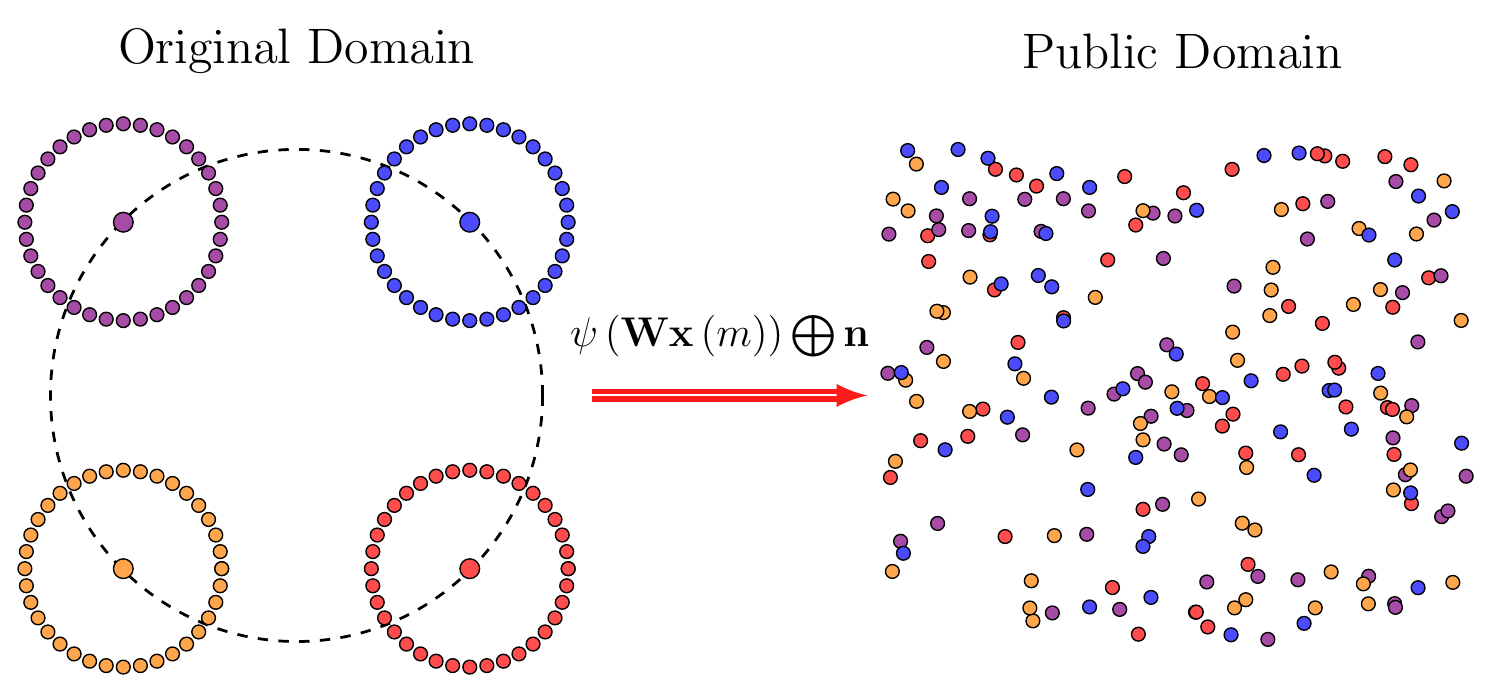}
\vspace{-5pt}
\caption{Schematic visualization of ambiguization on correlated clustered data used for Fig.~\ref{Fig:PDF}.}
 \label{Fig:VisuAmbiNoise4Clus}
    \vspace{-21pt}
\end{figure}

In \cite{mathon2013secure} the authors listed potential threats under the `honest but curious' model. According to our previous discussions the threat $S_1$ in \cite{mathon2013secure} precluded. In order to address threat $S_2$. i.e., clustering the database vectors from $\mathbf{a}\left(m \right), \forall m \in \left[ M \right]$, we generate four $512$-dimensional i.i.d. vectors with distribution $\mathcal{N} \left( \mathbf{0}, \mathbf{1} \right)$ and $1000$ $512$-dimensional i.i.d. vectors with distribution $\mathcal{N} \left( \mathbf{0}, \mathbf{0.1} \right)$. Then we add each $250$ (out of $1000$) low variance vectors to the four high variance ones (schematically shown in Fig.~\ref{Fig:VisuAmbiNoise4Clus}). This results in a database of $1000$ vectors adhering to four clusters. By defining sparsity ratios $\alpha_x = S_x / L $ and $\beta_x = S_{n_s} / L$, and denoting by $P_{\mathrm{intra}}$ and $P_{\mathrm{inter}}$ as probability density functions (PDFs) of `intra-cluster' and `inter-cluster' of distances, respectively, we intend to introduce a \textit{Kullback-Leibler divergence privacy leak measure}. To this end, we define distribution $P_1$ as mixture of $P_{\mathrm{intra}}$ and $P_{\mathrm{inter}}$, that is, $P_1 = \alpha_x \, P_{\mathrm{intra}} + \left( 1 - \alpha_x \right) P_{\mathrm{inter}}$, $0 \leq \alpha_x \leq 1$. Also, we denote the Gaussian distribution $\mathcal{N} \left( \mu_2 , \sigma_2^2 \right)$ by $P_2$, such that $\mu_2$ and $\sigma_2^2$ are the mean and variance of $P_1$, respectively. Therefore, the privacy leak measure can be defined by the Kullback-Leibler divergence (KLD) as $D \left( P_1 \| P_2 \right) = \alpha_x \, D \left( P_{\mathrm{intra}} \| P_2 \right) + \left( 1- \alpha_x  \right) D \left( P_{\mathrm{inter}} \| P_2 \right)$, where $D\left( P_1 \| P_2 \right) = \mathbb{E}_{P_1} \left[ \log \frac{P_1}{P_2} \right]$. Now, our privacy leak constraint can be expressed as $D \left( P_1 \| P_2 \right) \leq \delta$, where $\delta$ determines the allowable privacy leakage from database clustering. If all inter-class and intra-class distances follow the same distribution (close to Gaussian for large $N$ or $L$), then no clustering algorithm can differentiate them on the server side. In the same way, the  KLD $D \left( P_1 \| P_2 \right)$ will tend to zero. 

\vspace{-4pt}

In Fig.~\ref{Fig:PDF}, we illustrate the estimated PDFs of pairwise distances in the transform domain. The `solid lines' indicate $P_{\mathrm{intra}}$ and $P_{\mathrm{inter}}$ with different sparsity ratio $\beta_x$. As evident, by imposing ambiguization noise, the distributions $P_{\mathrm{intra}}$ and $P_{\mathrm{inter}}$ become unimodal Gaussian, therefore, the curious server cannot cluster database. Also, it is shown that by increasing the sparsity ratio of approximation, $\alpha_x$, we have to increase the sparsity ratio of ambiguization, $\beta_x$, in order to keep our privacy protection. The `dashed lines' indicate the corresponding Gaussian distribution fit to each `solid' plot. These figures clarify our main intuition of defining the KL divergence as a privacy protection measure in the ambiguized transform domain. Fig.~\ref{Fig:VisuAmbiNoise4Clus} visualizes how our model behaves in practice. Another interesting result is that $k$-means clustering will completely fail at the server, provided $D \left( P_1 \| P_2 \right) \leq \delta$. If the curious server runs the $k$-means scheme, the expected probability of correct clustering will be $\frac{1}{k}$, which shows maximal achievable Shannon entropy for random guessing. In other words, the bin (label) assignment at the server is completely random and server cannot learn the structure of database.

%
\begin{table}[!b]
\vspace{-14pt}
\centering
\caption{The upper limit of average distances between clusters when a user sends the positions to the server.}
\vspace{-5pt}
\label{Tab:FourClusterPositionServerSide}
{\renewcommand{\arraystretch}{1.15}%
\resizebox{\linewidth}{!}{%
\begin{tabular}{c|cccc|cccc|cccc|}
 \cline{2-13} 
\multicolumn{1}{c}{ } & \multicolumn{4}{c|}{$\beta_y \approx 0.04$}  &  \multicolumn{4}{c|}{ $\beta_y \approx 0.16$} &  \multicolumn{4}{c}{ $\beta_y \approx 1 - \alpha_y$}\cr
\hline
 \multirow{4}{*}{\begin{turn}{90}$\beta_x \approx 0.04$\end{turn}} & \cellcolor{gray!70}{0.9291} & \cellcolor{gray!10}{4.3568}    &  \cellcolor{gray!10}{4.1459}  & \cellcolor{gray!10}{4.0621} & \cellcolor{gray!55}{1.8378}    &  \cellcolor{gray!10}{4.6563}  &   \cellcolor{gray!10}{4.4239}  &  \cellcolor{gray!10}{4.3569}  &    \cellcolor{gray!40}{4.4524}     &   \cellcolor{gray!10}{6.3278}  &    \cellcolor{gray!10}{5.9947}  &   \cellcolor{gray!10}{5.9989}   \\
& \cellcolor{gray!10}{4.2247} &  \cellcolor{gray!70}{0.9042}    &   \cellcolor{gray!10}{4.2149}  &  \cellcolor{gray!10}{4.2231}   &  \cellcolor{gray!10}{4.5523} &  \cellcolor{gray!55}{1.8273}    &   \cellcolor{gray!10}{4.5442}  &  \cellcolor{gray!10}{4.5446}   &       \cellcolor{gray!10}{6.3276} &       \cellcolor{gray!40}{4.4522}   &   \cellcolor{gray!10}{6.3211} &   \cellcolor{gray!10}{6.3193} \\
& \cellcolor{gray!10}{4.1164} &  \cellcolor{gray!10}{4.3912}    &   \cellcolor{gray!70}{0.9278}  &  \cellcolor{gray!10}{4.1577}   &  \cellcolor{gray!10}{4.3965} &   \cellcolor{gray!10}{4.6747} &   \cellcolor{gray!55}{1.8319}    &  \cellcolor{gray!10}{4.4322}   &     \cellcolor{gray!10}{5.9947}   &  \cellcolor{gray!10}{6.3211}  &    \cellcolor{gray!40}{4.4513}     &    \cellcolor{gray!10}{6.0044} \\
& \cellcolor{gray!10}{4.0754} &  \cellcolor{gray!10}{4.4043}    &   \cellcolor{gray!10}{4.1618}  &  \cellcolor{gray!70}{0.9238}   &   \cellcolor{gray!10}{4.3832} &   \cellcolor{gray!10}{4.7061} &    \cellcolor{gray!10}{4.4547} &   \cellcolor{gray!55}{1.8776}    &       \cellcolor{gray!10}{5.9989}         &   \cellcolor{gray!10}{6.3193}  &    \cellcolor{gray!10}{6.0044} &    \cellcolor{gray!40}{4.4505}\\
\hline  
 \multirow{4}{*}{\begin{turn}{90} \tiny $\beta_x \approx 1 \! - \! \alpha_x$\end{turn}} & \cellcolor{gray!40}{5.4068} & \cellcolor{gray!10}{7.0727}    &  \cellcolor{gray!10}{6.7268}  & \cellcolor{gray!10}{6.6932} & \cellcolor{gray!30}{9.4215}    &  \cellcolor{gray!10}{10.4913}  &   \cellcolor{gray!10}{10.2535}  &  \cellcolor{gray!10}{10.2365}  &    \cellcolor{gray!10}{22.0792}     &   \cellcolor{gray!10}{22.6286}  &    \cellcolor{gray!10}{22.5188}  &   \cellcolor{gray!10}{22.5135}   \\
&  \cellcolor{gray!10}{7.0680} &  \cellcolor{gray!40}{5.4074}    &   \cellcolor{gray!10}{7.0168}  &  \cellcolor{gray!10}{6.9889}   &  \cellcolor{gray!10}{10.4819} &  \cellcolor{gray!30}{9.4254}    &   \cellcolor{gray!10}{10.4600}  &  \cellcolor{gray!10}{10.4372}   &       \cellcolor{gray!10}{22.6286} &       \cellcolor{gray!10}{22.0882}   &   \cellcolor{gray!10}{22.6180} &   \cellcolor{gray!10}{22.5998} \\
 & \cellcolor{gray!10}{6.7350} &  \cellcolor{gray!10}{7.0452}    &   \cellcolor{gray!40}{5.4027}  &  \cellcolor{gray!10}{6.7650}   &  \cellcolor{gray!10}{10.2609} &   \cellcolor{gray!10}{10.4687} &   \cellcolor{gray!30}{9.4181}    &  \cellcolor{gray!10}{10.2751}   &     \cellcolor{gray!10}{22.5188}   &  \cellcolor{gray!10}{22.6180}  &    \cellcolor{gray!10}{22.0790}     &    \cellcolor{gray!10}{22.5328} \\
& \cellcolor{gray!10}{6.6914} &  \cellcolor{gray!10}{7.0028}    &   \cellcolor{gray!10}{6.7587}  &  \cellcolor{gray!40}{5.4021}   &   \cellcolor{gray!10}{10.2311} &   \cellcolor{gray!10}{10.4405} &    \cellcolor{gray!10}{10.2758} &   \cellcolor{gray!30}{9.4142}    &       \cellcolor{gray!10}{22.5135}         &   \cellcolor{gray!10}{22.5998}  &    \cellcolor{gray!10}{22.5328} &    \cellcolor{gray!10}{22.0735}\cr
\hline
\end{tabular}%
}%
}%
\end{table}

\vspace{-3pt}

\begin{figure}[!t]
\centering
\includegraphics[width=5.4cm, height=2.82cm]{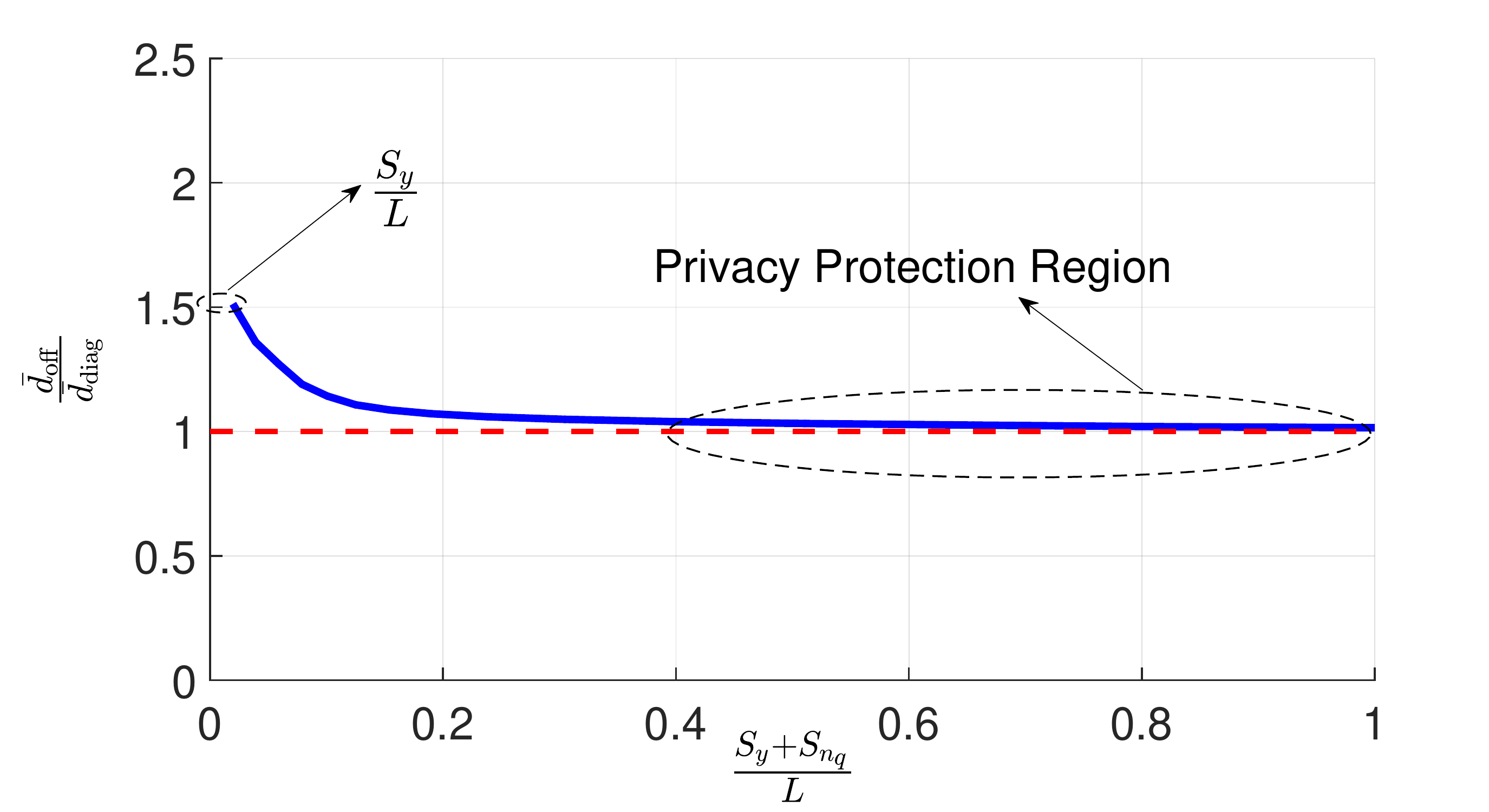}
\vspace{-5pt}
\caption{Privacy protection of database when probe discloses the positions.}
\vspace{-21pt}
 \label{Fig:PrivacyProtection}
\end{figure}

Although in our framework we design the decoder to be in the private domain, in order to address threats $S_3$ and $S_4$ of \cite{mathon2013secure} we generalize our framework and consider the \textit{decoder} in the \textit{public domain}. Now, we briefly discuss our solution for this scenario. In the public decoder scenario the data user (client side) just sends the positions of interest to the server and server returns lists of the vectors that have non-zero components for each position. We denote by $\mathcal{I} \subset \left[L\right]$ the set of indices that the user sends to the server such that $\mathrm{card} \left( \mathcal{I} \right) = S_y + S_{n_q}$, where $S_{n_q}$ is the number of ambiguized positions sent in addition to the $S_y$ interested positions. For request position $l \in \mathcal{I}$, the server returns $\mathcal{L}^+ \left(l \right) = \left\{ m \in \left[ M\right] : a_l \left(m\right) = +1 \right\}$ and $\mathcal{L}^- \left(l \right) = \left\{ m \in \left[ M\right] : a_l \left(m\right) = -1 \right\}$, where $\mathrm{card}\left( \mathcal{L}^+ \left(l \right) \right) \simeq \mathrm{card}\left( \mathcal{L}^- \left(l \right) \right) \simeq 0.5 M \left( \alpha_x + \beta_x \right) $. 
So, for a database consisting of $M = {10}^{9}$ vectors with parameters $S_y + S_{n_q} = 16$ and $\alpha_x + \beta_x \approx 0.03125$, we need $M \left( \alpha_x + \beta_x \right) \left( S_y + S_{n_q} \right)  \approx 62~\mathrm{MB}$ in order send back the lists  to the user. To avoid the communication overhead, the server can communicate only the data for a short list after score aggregation. The data user refines the list by rejecting the results for ambiguized positions. 
To investigate the upper limit on clustering based on k-means for the example above, we investigate the ratio between the inter-class and intra-class distances for various parameters. Assuming $L = 256$, $S_x = 10$ and $\sigma^2_{\mathbf{z}} = 0.15$, we provide the average distances in TABLE \ref{Tab:FourClusterPositionServerSide}, where $\alpha_y = S_y / L$ and $\beta_y = S_{n_q} / L$. 
Defining $\bar{d}_{\mathrm{diag}}$ as the average of diagonal entries and $\bar{d}_{\mathrm{off}}$ as the average of off-diagonal entries for pairwise average distance matrices shown in TABLE \ref{Tab:FourClusterPositionServerSide} and setting $S_{n_s} = L - S_x$, we plot the ratio of  $\bar{d}_{\mathrm{off}}$ to $\bar{d}_{\mathrm{diag}}$ as a function of $\left( S_y + S_{n_q} \right) / L = \alpha_y + \beta_y$ in Fig.~\ref{Fig:PrivacyProtection}. As evident, by increasing $S_{n_q}$ from $0$ to $L  - S_y$, the ratio  $\bar{d}_{\mathrm{off}} / \bar{d}_{\mathrm{diag}}$ converges to $1$, i.e., we reach maximal privacy when the query discloses the positions to the server.

\vspace{-10pt}


\section{Conclusion}\label{Sec:V}
\vspace{-4pt}

We have proposed a novel approach for privacy preserving identification based on a STC representation with ambiguization noise. Our scheme preserves distances up to a radius, as determined by the sparsity level of the approximation. However, for distances greater than our desired radius, the Euclidean distance between vectors in the transform domain becomes independent of the distance between the vectors in the original domain. Our method is generic, i.e., we have imposed no restrictions on the input data. One of the main points illustrated by this study is that we can share our database in the public domain, while preserving data privacy and even ensure the reconstruction of data for authorized users in contrast to known methods. The results show that the curious server cannot cluster the samples in the database, provided the Kullback-Leibler privacy leak constraint is satisfied. Furthermore, we proposed a public decoding method by introducing a novel formulation of query mechanism based on sending non-zero positions along with the ambiguous positions to server.

\vspace{-10pt}


\section*{Acknowledgment}
\vspace{-6pt}
B. Razeghi has been supported by the ERA-Net project ID\_IoT No 20CH21\_167534 and O. Taran by the SNF project No 200021\_165672. 

\vspace{-12pt}

\bibliographystyle{IEEEtran}
\bibliography{references}

\end{document}